\documentclass{JHEP3}
\usepackage{amsmath,amssymb,graphics,epsfig,subfigure}

\title{Bulk Matter Fields on a GRS-Inspired Braneworld}
\author{Yu-Xiao Liu,
        Chun-E Fu\footnote{corresponding author},
        Heng Guo,
        Shao-Wen Wei,
        Zhen-Hua Zhao,\\

 Institute of Theoretical Physics,
  Lanzhou University, Lanzhou 730000, P. R. China\\
  E-mail: \email{liuyx@lzu.edu.cn},
          \email{fuche08@lzu.cn},
          \email{guoh06@lzu.cn},
          \email{weishaow06@lzu.cn},
          \email{zhaozhh09@lzu.cn} }

\abstract{In this paper we investigate the localization and mass spectra of bulk matter fields on a Gergory-Rubakov-Sibiryakov-inspired braneworld. In this braneworld model, there are one thick brane located at the origin of the extra dimension and two thin branes at two sides. For spin $1/2$ fermions coupled with the background scalar $\phi$ via $\eta\bar{\Psi}\phi^p\Psi$ with $p$ a positive odd integer, the zero mode of left-hand fermions can be localized on the thick brane for finite distance of the two thin branes, and there exist some massive bound modes and resonance modes. The resonances correspond to the quasi-localized massive fermions. For free massless spin 0 scalars, the zero mode can not be localized on the thick brane when the two thin branes are located finitely. While for a massive scalar $\Phi$ coupled with itself and the background scalar field $\phi$, in order to get a localized zero mode on the thick brane, a fine-tuning relation should be introduced. Some massive bound modes and resonances also will appear. For spin 1 vectors, there is no bound KK mode because the effective potential felt by vectors vanishes outside the two thin branes. We also investigate the physics when the distance of the two thin branes tends to infinity.}


\begin{document}

\section{Introduction}

  Recently, the braneworld theory has attracted more and more
attention \cite{Rubakov1983,Akama1983,Antoniadis1990,ADD,rs,Lykken}. In this
theory, our matter fields are localized on the 3-dimensional brane, which is embedded in a higher dimensional space-time, while gravity is free to propagate in the whole space-time. The most important application of this theory is that it can solve some particle problems, such as mass hierarchy problem and cosmological
constant problem.

Two famous brane models brought forward  in early times are the
Arkani-Hamed-Dimopoulos-Dvali (ADD) large extra dimension brane
model \cite{ADD} and the Randall-Sundrum (RS) warped brane model
\cite{rs}. The extra dimensions of the ADD model are flat and compact, and the ADD model predicts the deviations from the 4D Newton law at sub-millimeter distances. While the RS model has an infinite extra dimension using a non-factorizable "warped" geometry. And in the RS model, there is a massless bound state graviton which should be interpreted as the usual 4D graviton. However, they are both thin branes, which have no thickness.

On the base of the RS model, another model called the
Gergory-Rubakov-Sibiryakov (GRS) model was proposed \cite{GRS}, in order to give some modifications of gravity at ultralarge scales. There are three thin branes in this model, which are located at $y=0$, $y=y_0$ and $y=-y_0$ (hereafter $y$ denotes the fifth coordinate), respectively. In the Ref. \cite{GRS2008}, the
GRS-inspired brane model was set up, in which the central thin brane
is replaced by a thick brane which is realized by a scalar field,
just as the usual thick branes
\cite{dewolfe,GremmPLB2000,gremm,Csaki,CamposPRL2002,varios,Guerrero2002,
ThickBraneDzhunushaliev,ThickBraneBazeia,ShtanovJCAP2009,Bazeia0808.2199}.

There are also other braneworld models, but for all models, an
important problem is the localization of various matter fields on the
brane \cite{M. O. Tahim08,0912.1029,B. Bajc0302069,Angel M.
Uranga0208014,Ichiro Oda0012013}. It has been known that massless
scalar fields and graviton can be localized on branes of different
types \cite{rs,BajcPLB2000}. But spin 1 Abelian vector fields can
not be localized on the RS brane in five dimensions, however, it can
be localized on the RS brane in some higher-dimensional cases
\cite{OdaPLB2000113}, or on the thick de Sitter brane and the Weyl
thick brane \cite{LiuJCAP2009,Liu0803}.

The localization of fermions is more interesting \cite{Liuzhao1004,C. A. S.
Almeida0901.3543,Liu0907.0910,Alejandra0601161,Ratna0806.0455} and
has been extensively analyzed in many papers, such as with gauge
field background \cite{Parameswaran0608074,LiuJHEP2007}, with
supergravity background \cite{Mario}, with vortex background
\cite{LiuNPB2007,LiuVortexFermion,Rafael200803,StojkovicPRD}. It was
also found that there may exist a single bound state and a
continuous gapless spectrum of massive fermion Kaluza--Klein (KK)
states
\cite{ThickBrane1,ThickBrane2,ThickBrane3,Liu0708,Liu0709,20082009},
or finite discrete KK states (mass gaps) and a continuous
spectrum starting at a positive $m^2$
\cite{Liu0803,ThickBrane4,0803.1458} with the scalar-fermion
coupling. Some resonances also exist
\cite{KoleyCQG2005,YuXiaoLiu0909.2312}, whose life-time depends on
the parameters of the scalar potential.

In this paper we investigate the localization and mass spectra
of bulk matter fields on the GRS-inspired braneworld. Our paper is
organized as follows: In Sec. \ref{SecModel}, we first give a brief
review of the GRS-inspired braneworld.
Then, in Sec. \ref{SecLocalize}, we study the localization and
mass spectra of spin 1/2 fermions, spin 0 scalars and spin 1
vectors on the GRS-inspired braneworld by presenting the shapes of
the potentials of the corresponding Schr\"{o}dinger equations.
Finally, in Sec. \ref{secConclusion}, we make a brief discussion and a
conclusion.

\section{Review of the GRS-inspired model}
\label{SecModel}

  An interesting model which has the feature of predicting modifications
of gravity at ultralarge scales is the one proposed by Gregory,
Rubakov and Sibiryakov (GRS). In the usual GRS model, considering the fifth dimension $y$ to be infinite and non-compact, there are three thin branes, one is at $y=0$ with positive tension $\sigma$, the other two are at $y=y_0$ and $y=-y_0$, respectively, both with the tension $-\sigma/2$. Note that the total charge in this braneworld is zero. The cosmological constant $\Lambda$ between the positive and the negative tension branes is assumed to be negative, namely, the space for $|y|\leq y_0$ is $AdS_5$, which is the maximally symmetric space. Beyond $|y_0|$ the space is the Minkowski space.  The system is assumed to have orbifold symmetry. So we can deal with the domain $y>0$ only, because for negative values of $y$ the system is just a mirror reflection.

If the central thin brane is replaced by a thick brane realized by a suitable configuration of a bulk scalar field \cite{scalarGRS}, which is called the GRS-inspired brane \cite{GRS2008}, the action of this model is
\begin{eqnarray}
 S = \int d^5 x \sqrt{-G} \left[ 2M^3 R - \frac{1}{2}(\nabla \phi)^2 - V(\phi)-\Lambda \right]
       -\int d^4 x \sqrt{-g} \sigma(\phi),
\label{action}
\end{eqnarray}
where $M$ is the basic Planck energy scale in the 5-dimensional
space-time, $G_{M N}$ denotes the 5-dimensional metric and $g_{\mu
\nu}$ the 4-dimensional one. The first part of the action
contains the gravitation and the scalar field, while the second part
is the contribution of a thin brane. {{ The background scalar $\phi$ is considered to be only the function of the extra dimensional coordinate for easier, i.e., $\phi=\phi(y)$, and the brane tension depends on
the scalar field only, i.e., $\sigma=\sigma(\phi)$, because the branes are treated as the cross-sections of the bulk.}}


{{ In general, the geometry of a 4-dimensional brane with maximal symmetry can be supposed as $M_4$, $dS_4$ or $AdS_4$. Here, we consider the simplest case, the $M_4$ brane. Hence, the metric of the background space-time is assumed as \cite{AdS5} }}
\begin{eqnarray}
 ds^2&=& \text{e}^{2A(y)}\eta_{\mu\nu}dx^\mu dx^\nu+dy^2,\label{linee1}
\end{eqnarray}
where $\text{e}^{2A(y)}$ is the warp factor. For convenience, the metric (\ref{linee1}) can be rewritten as
\begin{eqnarray}
 ds^2&=&\text{e}^{2A(z)}\big(\eta_{\mu\nu}dx^\mu dx^\nu
          + dz^2\big), \label{linee}
\end{eqnarray}
where the two coordinate systems are connected by
$dy=\text{e}^{A(z)} dz$, so the point $y=0$ is mapped to
$z=0$, while $y_0$ to $z_0$. And the extra dimension $z$ is also infinite.

 Then the equations of motion generated from
the action (\ref{action}) are
\begin{eqnarray}
R_{M N}-\frac{1}{2}G_{M N}R&=&\frac{1}{4M^3}\bigg[\nabla_M \phi \nabla_N \phi -
                           G_{M N}\big(\frac{1}{2} (\nabla \phi)^2 + V(\phi) \big) \nonumber \\
                           & & -G_{M N}\Lambda - \sigma \delta(z - z_0) \text{e}^{-A(z_0)} g_{\mu \nu}\delta^\mu_M \delta^\nu_N \bigg],\\
\nabla^2 \phi +\frac{d V}{d \phi}  &=& 0,
\end{eqnarray}
which can be reduced to the following coupled nonlinear
differential equations with the ansatz (\ref{linee}):
\begin{eqnarray}
 \phi'^2 & = & 12M^3(A'^2-A'') - \sigma \delta(z-z_0)\text{e}^{-A(z_0)}, \\
 V(\phi) & = & 6M^3(-3A'^2-A'') \text{e}^{-2A} -\Lambda
 -\frac{\sigma}{2}\delta(z-z_0)\text{e}^{-A(z_0)},\\
 \frac{dV(\phi)}{d\phi}&=&(3A'\phi'+\phi'')\text{e}^{-2A}
                      -\frac{d\sigma(\phi)}{d\phi}\delta(z-z_0)\text{e}^{-A(z_0)},
\end{eqnarray}
where the prime stands for the derivative with respect to $z$. Note that, here we only consider the range of $z\geq 0$, so terms containing $\delta(z+z_0)$ do not appear in above equations.

The background solution for this GRS-inspired domain wall is \cite{GRS2008} \footnote{The potential $V$ in (\ref{Va}) for $z\geq z_0$ is a little different from that in Ref. \cite{GRS2008}, where $V= {\frac{6 k^2 M^3}{b}}\bigg(\frac{4 \tan^2 (\frac{\phi}{\sqrt{12}M^{3/2}}) - 1 }{\tan^2 (\frac{\phi}{\sqrt{12}M^{3/2}}) + 1}
 \bigg)- \Lambda$.}
\begin{eqnarray}
 A   &=& \frac{1}{2} \ln\frac{b}{1+(k z)^2},\label{warpfactor1}\\
 \phi&=& \sqrt{12} M^{3/2} \arctan(k z),\\
 V   &=& -{\frac{6 k^2 M^3}{b}}\bigg(\frac{4 \tan^2 (\frac{\phi}{\sqrt{12}M^{3/2}}) - 1 }{\tan^2 (\frac{\phi}{\sqrt{12}M^{3/2}}) + 1}\bigg)-\Lambda, \label{Va}
\end{eqnarray}
for $|z| < z_0$, and
\begin{eqnarray}
 A   &=& \frac{1}{2} \ln\frac{b}{1+(k z_0)^2},\label{warpfactor2}\\
 \phi&=& \sqrt{12} M^{3/2} \arctan(k z_0),\\
 V   &=& -\Lambda,
\end{eqnarray}
for $|z|>z_0$, where $k$ and $b$ are both constants and $k$ plays the role of the central wall's thickness. With the normalization condition $e^{2A(0)}=1$ for the warp factor, the parameter $b$ can be set to 1. And with the solution of the warp factor above we can calculate the exact relation between the two coordinates:
\begin{eqnarray}
 y=\int_0^z\text{e}^{A(\bar{z})}d\bar{z}=\frac{1}{k}\text{arcsinh}(kz),
\end{eqnarray}
for $-z_0<z<z_0$, and
\begin{eqnarray}
 y=\int_0^z\text{e}^{A(\bar{z})}d\bar{z}=c+\frac{1}{\sqrt{1+k^2z_0^2}}z,
\end{eqnarray}
for $|z|>z_0$, where $c$ is a constant as $c=\frac{1}{k}\text{arcsinh}(kz_0)+\frac{z_0}{\sqrt{1+k^2z_0^2}}$. So it is clear the extra coordinate $z$ is also infinite.

It can be seen that $V(\phi)$ may be discontinuous at $z_0$, unless we set $z_0=1/(2k)$. However, the localizing potential of gravity for $|z|< z_0$, i.e., $V_l=\frac{3 k^2 [5(kz)^2-2]}{4 [1 + (kz)^2]^2}$ \cite{GRS2008}, can not reach zero if we set $z_0=\frac{1}{2k}$ because the zero point of $V_l$, $z_1=\sqrt{\frac{2}{5}}\frac{1}{k}$, is larger than $\frac{1}{2k}$. So, considering that $V_l=0$ for $|z|> z_0$ (this is because we have a Minkowski space for $|z|>z_0$), we would get a localizing potential with no barrier if we perform the patch. The detailed discussion can be found in Ref \cite{GRS2008}.

\begin{figure*}[htb]
\begin{center}
\includegraphics[width=7.5cm]{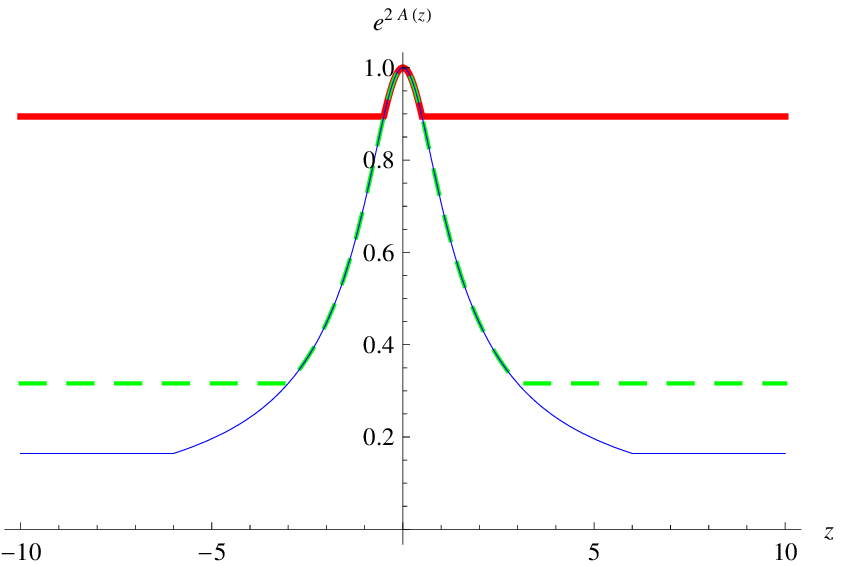}
\includegraphics[width=7.5cm]{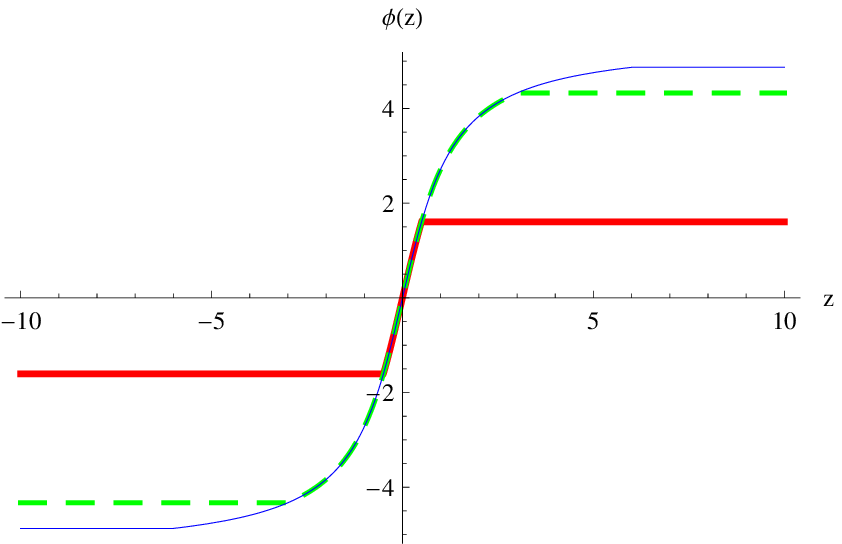}
\end{center}
 \caption{(Color online) The shapes of the warp factor
$\text{e}^{2A(z)}$ and the kink $\phi(z)$. The parameters are set to
$k=1$, $z_0=0.5$ for red thick line { (which is the case $2kz_0=1$)}, $z_0=3$ for green dashing line, and $z_0=6$ for blue thin line.}
 \label{figWarpFactorkink}
\end{figure*}

\begin{figure*}[htb]
\begin{center}
\subfigure[{{ $k=0.1$}}]{\label{figEnergyDendity_a}
\includegraphics[width=6.5cm]{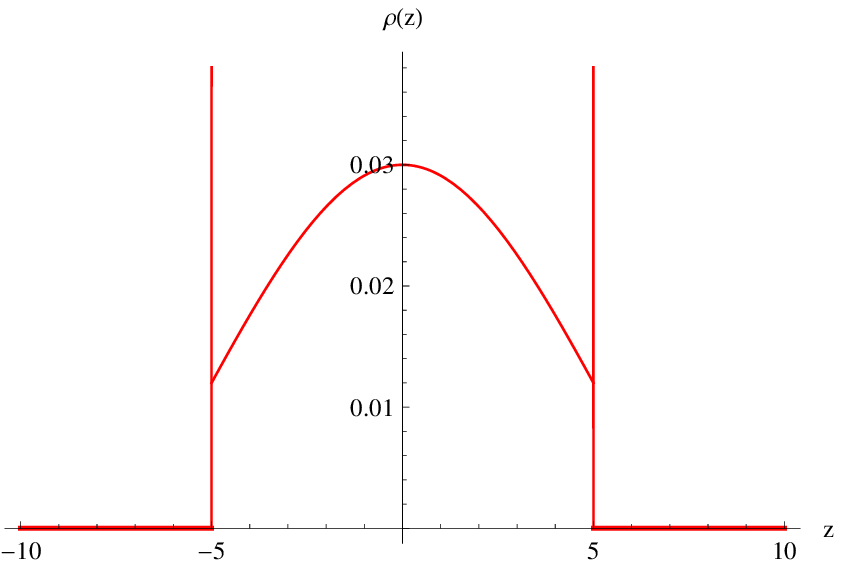}}
\subfigure[$k=3$]{\label{figEnergyDendity_b}
\includegraphics[width=6.5cm]{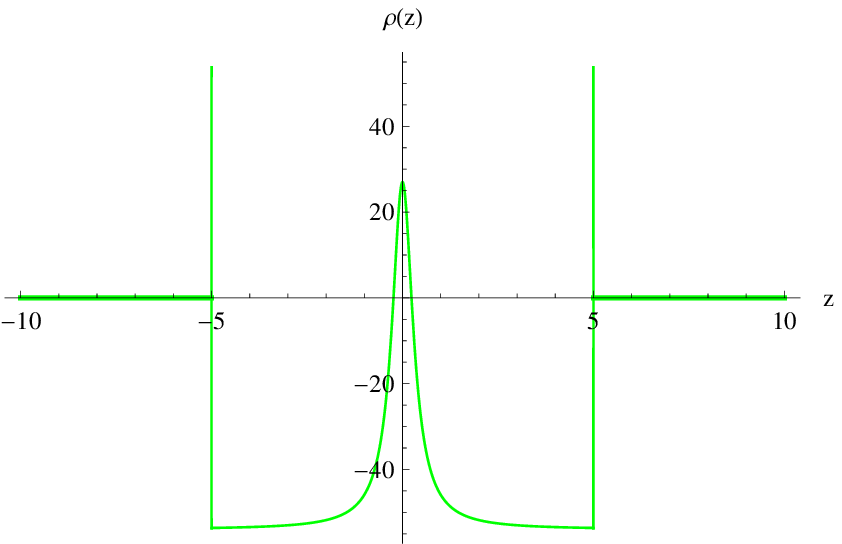}}
\end{center}
\caption{(Color online) The shapes of the energy density $\rho(z)$. The
parameters are set to $z_0=5$, {{ $k=0.1$}} for the left figure {{ (which is the case $2kz_0=1$)}}, and $k=3$
for the right figure.}
 \label{figEnergyDendity}
\end{figure*}

The scalar configuration is in fact a kink. It is clear that beyond $z_0$ the warp factor and the scalar field are both constants. So the expression of the energy density $\rho$ is
\begin{equation}
 \rho=\left\{\begin{array}{ll}
        \frac{12 k^2 (1- 2 k^2 z^2)}{1+k^2 z^2}, & |z| < z_0\\
        -\frac{\sigma}{2}(1+k^2 z_0^2)^{3/2} \big[\delta(z-z_0)+\delta(z+z_0)], & |z| > z_0
      \end{array}~~.\right.
\end{equation}
{{ The tension of the thin branes $\sigma$ satisfies $\sigma=-\frac{12 M^3 k^2 z_0}{\sqrt{1+k^2 z_0^2}}$, which is calculated from the junction condition at $z=\pm z_0$ \cite{GRS2008}. We can see that the energy density vanishes at $|z| > z_0$, which is because the space beyond $z_0$ is Minkowski.}} For convenience, we set $M=1$ in what follows. The shapes of the warp factor
$\text{e}^{2A(z)}$, the kink $\phi(z)$, and the energy density
$\rho(z)$ are shown in Fig. \ref{figWarpFactorkink} and Fig.
\ref{figEnergyDendity}.


\section{Localization and mass spectrum of various matter fields on the brane}
\label{SecLocalize}

  In this section, we will investigate the localization and mass
spectra of various bulk matter fields such as spin 1/2 fermions, spin 0
scalars and spin 1 vectors on the GRS-inspired braneworld by
presenting the potentials of the corresponding Schr\"{o}dinger
equations. We treat the bulk matter fields considered below as small perturbations around the background \cite{9808016,9905186}, namely, we neglect the back-reaction of bulk matter fields on the background geometry. So the bulk fields make little contribution to the bulk energy, and the background solutions given in previous section are still valid.


\subsection{Spin 1/2 fermion fields}

 Firstly, we will investigate the localization and mass spectra of the
spin $1/2$ fermions on the GRS-inspired braneworld by means of the
gravitational interaction and scalar-fermion coupling.

Using our usual set-up, we consider a massless spin $1/2$ fermion coupled with gravity and the background scalar $\phi$ in 5-dimensional
space, of which the Dirac action is
\begin{eqnarray}
 S_{1/2} &=& \int d^5 x \sqrt{-G} \bigg(\bar{\Psi} \Gamma^M (\partial_M + \omega_M) \Psi
       -\eta \bar{\Psi} F(\phi) \Psi\bigg),
\label{DiracAction}
\end{eqnarray}
where $\Gamma^M=(\text{e}^{-A}\gamma^{\mu},\text{e}^{-A}\gamma^5)$,
$\omega_\mu =\frac{1}{2}(\partial_{z}A) \gamma_\mu \gamma_5$,
$\omega_5 =0$, and $\eta$ is the coupling constant between the spinor
$\Psi$ and the scalar field $\phi$ with $F(\phi)$ the type of the
coupling. $\gamma^{\mu}$ and $\gamma^5$ are the usual flat gamma
matrices in the Dirac representation. The 5-dimensional Dirac
equation is read as
\begin{eqnarray}
 \left\{ \gamma^{\mu}\partial_{\mu}
         + \gamma^5 \left(\partial_z  +2 \partial_{z} A \right)
         -\eta\; \text{e}^A F(\phi)
 \right \} \Psi =0, \label{DiracEq1}
\end{eqnarray}
where $\gamma^{\mu} \partial_{\mu}$ is the 4-dimensional Dirac
operator.

Now we would like to decompose the 5-dimensional spinor $\Psi$ as
\begin{equation}
 \Psi(x,z) = \text{e}^{-2A}\sum_n\bigg(\psi_{Ln}(x) f_{Ln}(z)
 +\psi_{Rn}(x) f_{Rn}(z)\bigg),
\end{equation}
where $\psi_{Ln}=- \gamma^5 \psi_{Ln}$ and $\psi_{Rn}=\gamma^5 \psi_{Rn}$.
Substituting the above expression of $\Psi(x,z)$ into the equation of motion (\ref{DiracEq1}), and assuming that $\psi_{Ln,Rn}(x)$ satisfy the 4-dimensional
massive Dirac equations $\gamma^{\mu}\partial_{\mu}\psi_{Ln,Rn}(x)
=m_n\psi_{Rn,Ln}(x)$, one can
obtain the coupled equations of the fermion KK modes $f_{Ln}(z)$ and $f_{Rn}(z)$:
\begin{subequations}\label{CoupleEq1}
\begin{eqnarray}
 \left[\partial_z + \eta\;\text{e}^A F(\phi) \right]f_{Ln}(z)
  &=&  ~~m_n f_{Rn}(z), \label{CoupleEq1a}  \\
 \left[\partial_z- \eta\;\text{e}^A F(\phi) \right]f_{Rn}(z)
  &=&  -m_n f_{Ln}(z). \label{CoupleEq1b}
\end{eqnarray}
\end{subequations}
Then from the above coupled equations, we get the Schr\"{o}dinger-like equations for the KK modes of the left and right chiral fermions:
\begin{subequations}\label{SchEqFermion}
\begin{eqnarray}
  \big(-\partial^2_z + V_L(z) \big)f_{Ln}
            &=&m_n^2 f_{Ln},
   \label{SchEqLeftFermion}  \\
  \big(-\partial^2_z + V_R(z) \big)f_{Rn}
            &=&m_n^2 f_{Rn},
   \label{SchEqRightFermion}
\end{eqnarray}
\end{subequations}
where the effective potentials are
\begin{subequations}\label{Vfermion}
\begin{eqnarray}
  V_L(z)&=& \big(\eta\;\text{e}^{A}   F(\phi)\big)^2
     - \partial_z\big(\eta\;\text{e}^{A}   F(\phi)\big), \label{VL}\\
  V_R(z)&=&   V_L(z)|_{\eta \rightarrow -\eta}. \label{VR}
\end{eqnarray}
\end{subequations}
Moreover we need the following orthonormality conditions for
$f_{L{n}}$ and $f_{R{n}}$ to get the standard 4-dimensional action
for massive chiral fermions:
\begin{eqnarray}
&& \int_{-\infty}^{\infty} f_{Lm}(z) f_{Ln}(z)dz
  =\delta_{mn},\nonumber\\
&&\int_{-\infty}^{\infty} f_{Rm}(z) f_{Rn}(z)dz
  =\delta_{mn},\label{orthonormality}\\
&& \int_{-\infty}^{\infty} f_{Lm}(z) f_{Rn}(z)dz=0.
 \nonumber
\end{eqnarray}

  It can be seen that the potentials $V_L(z)$ and $V_R(z)$ depend on three factors,
i.e., the scalar-fermion coupling constant $\eta$, the warp factor
$\text{e}^{A(z)}$ and the type of scalar-fermion coupling $F(\phi)$.
In order to localize left- or right-hand fermions, the effective
potential $V_L(z)$ or $V_R(z)$ should have a minimum value at the
location of the thick brane. Furthermore, as we demand that
$V_{L,R}(z)$ are invariant under $Z_2$ reflection symmetry
$z\rightarrow -z$, $F(\phi(z))$ should be an odd function of $z$.
Since the kink configuration of the scalar $\phi(z)$ is $Z_2$-odd,
$F(\phi)$ should be an odd function of $\phi$. Thus we have
$F(\phi(0))=0$ and $V_L(0)=-V_R(0)=-\eta\partial_z F(\phi(0))$,
which would result in that only one of the massless left and right
chiral fermions can be localized on the brane.

From the solutions of the metric and the scalar field given in previous section, we
can see that, for any type of scalar-fermion coupling, when $|z| \geq
z_0$ the value of the potentials is a constant, i.e.,
$V_{L,R}(\pm\infty)=V_{L,R}(z_0)$. So there may exist discrete bound
KK modes for $m_n^2<V_{L,R}(z_0)$ and continuous ones for
$m_n^2>V_{L,R}(z_0)$. Here, for simplicity, we only consider the
simplest Yukawa coupling $F(\phi)=\phi$ and the generalized one
$F(\phi)=\phi^p$ with $p$ an odd positive integer.

For the simplest Yukawa coupling $F(\phi)=\phi$, the potentials take
the following form
\begin{subequations}\label{VLa}
\begin{eqnarray}
 V_L(z) &=& \left\{
    \begin{array}{ll}
      \frac{12 \eta^2 \arctan^{2}(k z)}{1 + k^2 z^2}
      +\frac{2\sqrt{3}\eta k}{(1 + k^2 z^2)^{3/2}}
      \big[ k z \arctan(k z)-1\big], & |z|<z_0    \\
      \frac{12\eta^2 \arctan^{2}(k z_0)}{1 + k^2 z_0^2}, & |z| \geq z_0 \\
    \end{array} \right., \\
 V_R(z) &=& V_L|_{\eta \rightarrow -\eta},
\end{eqnarray}\end{subequations}
and for the generalized Yukawa coupling with $p\geq3$, the potentials are
\begin{subequations}\label{VLb}
\begin{eqnarray}
 V_L(z) &=& \left\{
    \begin{array}{ll}
      \frac{12^p \eta^2 \arctan^{2p}(k z)}{1+k^2 z^2}
      +\frac{12^{p/2} \eta k}{(1 + k^2 z^2)^{3/2}}
      \arctan^{p-1}(k z)\big[k z\arctan(k z) -p \big], & |z|<z_0     \\
      \frac{12^p \eta^2 \arctan^{2p}(k z_0)}{1 + k^2 z_0^2}, & |z| \geq z_0 \\
    \end{array} \right., \\
 V_R(z) &=& V_L|_{\eta \rightarrow -\eta}.
\end{eqnarray}\end{subequations}

\begin{figure*}[htb]
\begin{center}
\includegraphics[width=7.5cm]{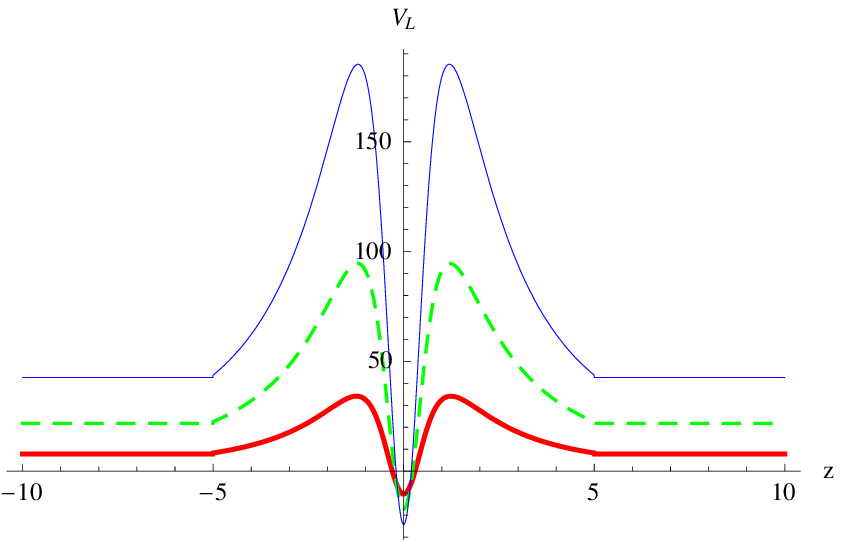}
\includegraphics[width=7.5cm]{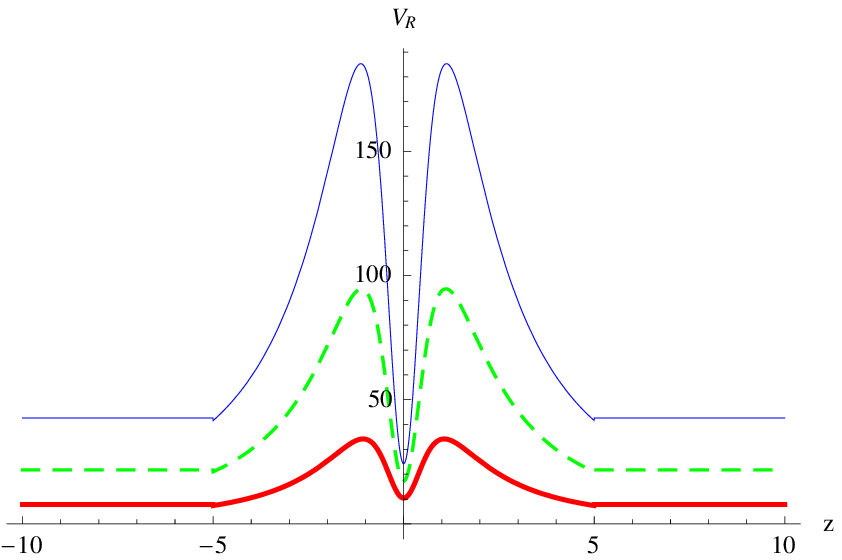}
\end{center}
 \caption{(Color online) The shapes of potentials $V_L$ (the left figure)
 and $V_R$ (the right figure) with $F(\phi)=\phi$
 for different coupling constant $\eta$.
 The parameters are set to $k=1,z_0=5$,
 $\eta=3$ for red thick lines, $\eta=5$ for green dashing lines,
 and $\eta=7$ for blue thin lines.}
 \label{figVLRa}
\end{figure*}

\begin{figure*}[htb]
\begin{center}
\includegraphics[width=7.5cm]{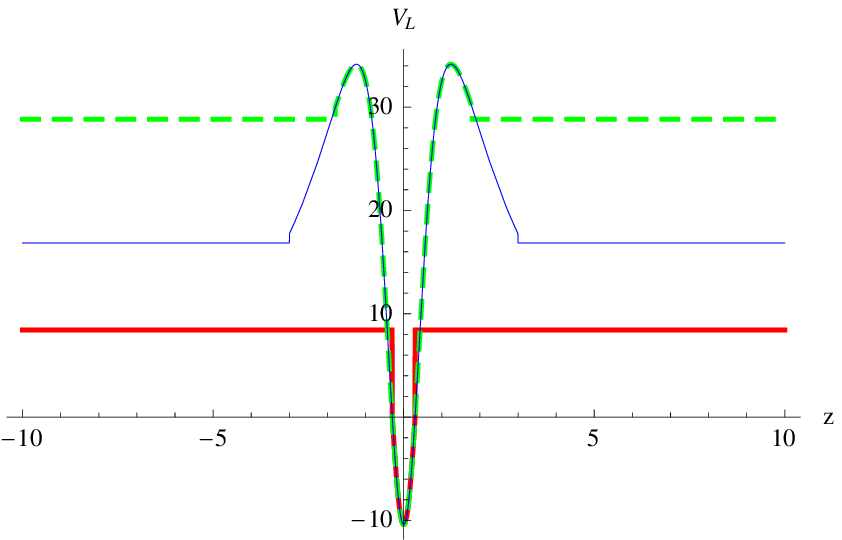}
\includegraphics[width=7.5cm]{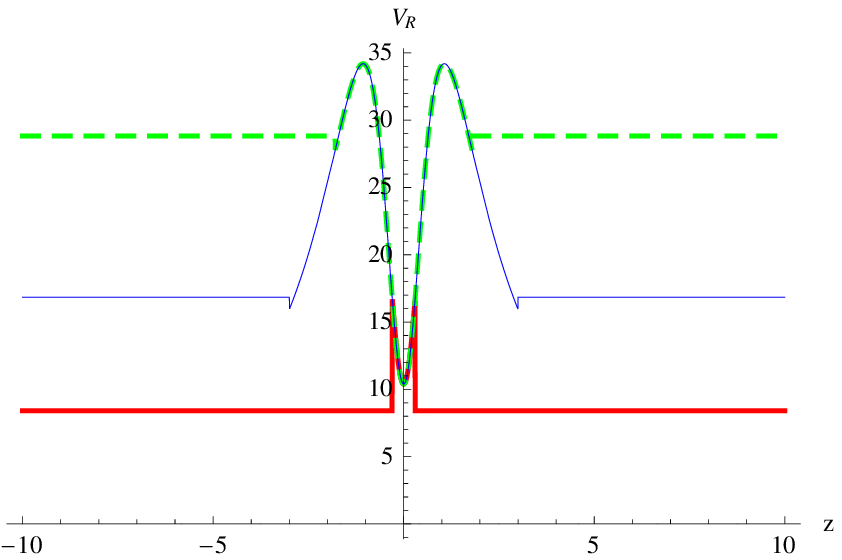}
\end{center}
 \caption{(Color online) The shapes of potentials $V_L$ (the left figure)
 and $V_R$ (the right figure) with $F(\phi)=\phi$ for different $z_0$.
 The parameters are set to $k=1,\eta=3$,
 $z_0=0.7$ for red lines, $z_0=2$ for green dashing lines,
 and $z_0=3$ for blue thin lines.}
 \label{figVLRb}
\end{figure*}

\begin{figure*}[htb]
\begin{center}
\subfigure[$p=3$]{\label{figVLRc_a}
\includegraphics[width=6.5cm]{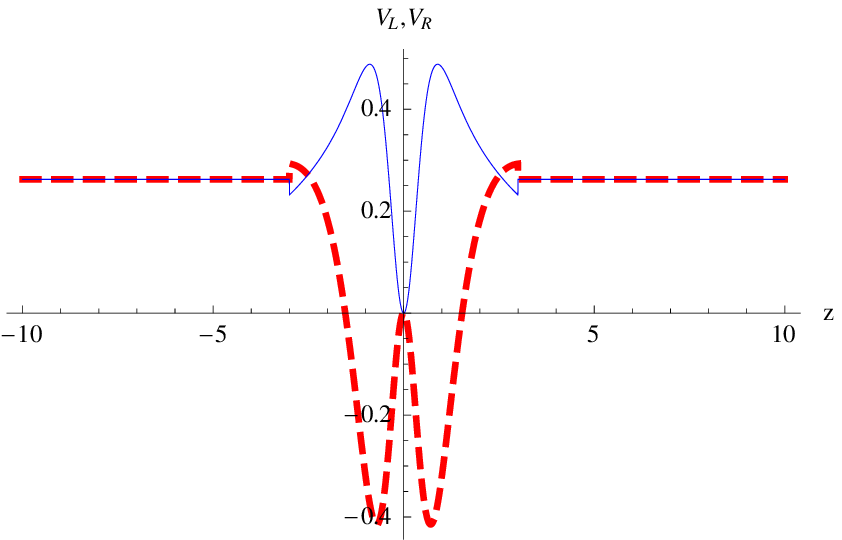}}
\subfigure[$p=5$]{\label{figVLRc_b}
\includegraphics[width=6.5cm]{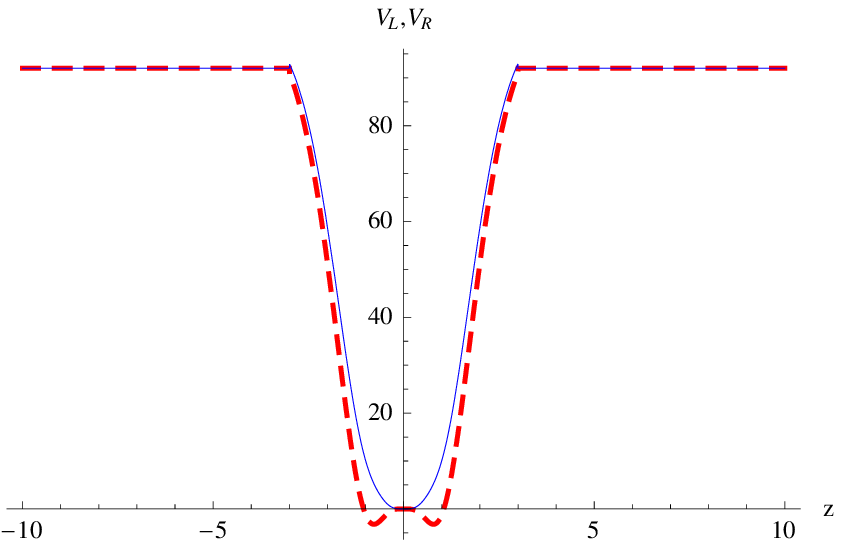}}
\end{center}
\caption{(Color online) The shapes of potentials $V_L$ (the red thick
dashing lines)
 and $V_R$ (the blue thin lines) with $F(\phi)=\phi^p$ for different $p$.
 The parameters are set to $k=1,~z_0=3,~\eta=0.02$,
 $p=3$ for the left figure, $p=5$ for the right figure.}
 \label{figVLRc}
\end{figure*}

The values of the potentials for left and right chiral fermions at
$z=0$ are
\begin{eqnarray}
 V_L(0) &=&-V_R(0) = - 2 \sqrt{3} k \eta, ~~~\text{for}~~~~ p=1,\\
 V_L(0) &=&-V_R(0) = 0, ~~~~~~~~~~~~~\text{for}~~~~p\geq3 ,
\end{eqnarray}
and $V_{L,R}(\pm\infty)=V_{L,R}(z_0)$ at $y\rightarrow\pm\infty$, so it is clear the potentials are P\"{o}schl-Teller-like potentials. For positive coupling constant, the minimum value of the potentials for left-hand fermions is negative and for right-hand fermions positive, so only left-hand fermions have one zero mode. While for massive KK modes, the mass spectra are the same for both chiral fermions, which can be seen from ({\ref{CoupleEq1a}}) ({\ref{CoupleEq1b}}). Because $V_{L,R}(z\geq z_0)=V_{L,R}(z_0)$ is a constant, so the value of the potentials at $z=z_0$ is very important for the number of the massive bound KK modes. The value of $V_{L,R}(z_0)$ is determined by three parameters: $z_0$, $\eta$ and $p$. It can be seen that that $V_{L,R}(z_0)$ increases with $\eta$ and $p$. And there is a special point of $z_0$, for which $V_{L,R}(z_0)$ takes its maximum (see Fig. \ref{figVLz0}). The shapes of the potentials for different parameters are plotted in Fig.~\ref{figVLRa}, Fig.~\ref{figVLRb} and Fig.~\ref{figVLRc}, with Fig.~\ref{figVLRa} and Fig.~\ref{figVLRb} for the simplest Yukawa coupling, and Fig.~\ref{figVLRc} for the generalized one.

\subsubsection{The zero mode}

  The zero mode for the left-hand fermions can be solved from
({\ref{CoupleEq1a}}) by setting $m_0=0$:
\begin{equation}
 f_{L0}(z) \propto \exp\left(-\eta\int^z_0 d\widetilde{z}
 \text{e}^{A(\widetilde{z})}\phi(\widetilde{z})\right).
  \label{zeroModefL0}
\end{equation}
In order to check the normalization condition (\ref{orthonormality})
for the zero mode (\ref{zeroModefL0}), we need to check whether the
integral
\begin{equation}
 \int f_{L0}^2(z) dz \propto
 \int \exp\left(-2\eta\int^z d\widetilde{z}
 \text{e}^{A(\widetilde{z})}\phi(\widetilde{z})\right) dz
 \label{condition1}
\end{equation}
is finite. For the integral $\int^z dz\text{e}^{A}\phi$, we only need
to consider the part of $|z|\geq z_0$, because the integral is
finite for the integrand range $-z_0{\leq} z{\leq} z_0$. Since
\begin{eqnarray}
 2\;\text{e}^{A} \phi
 &=& 2\sqrt{\frac{12}{1+(k z_0)^2}} \arctan(k z_0)
   \nonumber \equiv c ~~ \text{for} ~~|z| \geq z_0,
\end{eqnarray}
where $c$ is a positive constant for positive $k$, the integrand
$f_{L0}^2$ is
\begin{eqnarray}
  f_{L0}^2(z)\propto\text{e}^{-2 \eta c z}
            ~~~~~ \text{for} ~~~|z|\geq z_0,
\end{eqnarray}
which indicates that for a finite $z_0$ the integral (\ref{condition1}) is finite and the zero mode of left-hand fermions can be localized on the brane
without any other condition. However, for an infinite $z_0$, the integral $\int^z dz\text{e}^{A}\phi$ tends to:
\begin{eqnarray}
 \int^z dz\text{e}^{A}\phi \rightarrow \frac{\sqrt{3}\pi}{k}\ln z,~~~\text{for}~~~ y\rightarrow \pm\infty,
\end{eqnarray}
then the integrand $f_{L0}^2$ becomes
\begin{eqnarray}
   f_{L0}^2(z) \propto  z^{\frac{-2\sqrt{3}\eta\pi}{k}},
\end{eqnarray}
so only when $\eta>\frac{k}{2\sqrt{3}\pi}$, the zero mode for the left chiral fermion can be localized on the brane, which is corresponding to that in Ref \cite{Liu0907.0910}.

\subsubsection{The massive bound modes}

  About the massive KK modes for both left-hand and
right-hand fermions, as discussed before, the only difference between the mass spectra of left-hand and right-hand fermions is that there exists one zero mode for left-hand fermions but does not for right-hand ones for any positive coupling constant $\eta$. In the following discussions we will see the impact of different parameters on the KK modes through their impact on the potentials.

  From Fig.~\ref{figVLRa}, we can see that with the increase
of $\eta$, {{ the depth of the potential well for both chiral fermions increases,}} which means that more massive bound KK modes would appear. Using the numerical
method, we get the mass spectrum of the KK modes for both chiral fermions with $k=1,z_0=1$ and different $\eta$. For $\eta=1$, there is only the zero mode for
left-hand fermions. For $\eta=3$, there are two massive bound KK
modes for both chiral fermions. For $\eta=5$, there are
three ones. The mass spectra are
\begin{eqnarray}
\begin{array}{ll}
  m_{Ln}^2 =\{0, 18.19, 30.64\} \cup ~[33.01,\infty)
      & ~~~~\text{for} ~~~~\eta=3, \\
  m_{Rn}^2 =\{~~~18.19, 30.64\} \cup ~[33.01,\infty)
     & ~~~~\text{for} ~~~~\eta=3, \\
\end{array}
\label{spectraMSLn1CaseII}
\end{eqnarray}
\begin{equation}
\begin{array}{ll}
  m_{Ln}^2 =\{0, 32.09, 58.93, 79.97\} \cup ~[92.53,\infty)
      & ~~~\text{for} ~~~~\eta=5,\\
  m_{Rn}^2 =\{~~~32.09, 58.93, 79.97\} \cup ~[92.53,\infty)
      & ~~~\text{for} ~~~~\eta=5.\\
\end{array}
\label{spectraMSLn1CaseII}
\end{equation}

Next let us discuss the effect of the type of the scalar-fermion
coupling, i.e., the effect of $p$. From Fig.~\ref{figVLRc}, it can
be seen that with the increase of $p$, {{ the depth of the potential well for both chiral fermions increases sharply,}} which means that there would
be more bound KK modes. Here we calculate the mass spectrum for
$p=3$, $k=1$, $\eta=1$ and $z_0=1$:
\begin{eqnarray}
\begin{array}{ll}
  m_{Ln}^2 =\{0, 10.88, 33.47, 58.58, 86.42, 116.00,146.60, 177.44\} ~\cup ~[202.79,\infty),  \\
  m_{Rn}^2 =\{~~~10.88, 33.47, 58.58, 86.42, 116.00,146.60, 177.44\} ~\cup ~[202.79,\infty).\\
\end{array}
\label{spectraMSLn1CaseII}
\end{eqnarray}
Compared with the simplest Yukawa coupling (the case with $p=1$), it
is clear that when the coupling index $p$ increases, the potentials
could trap more bound KK modes.

\begin{figure*}[htb]
\begin{center}
\includegraphics[width=7.5cm]{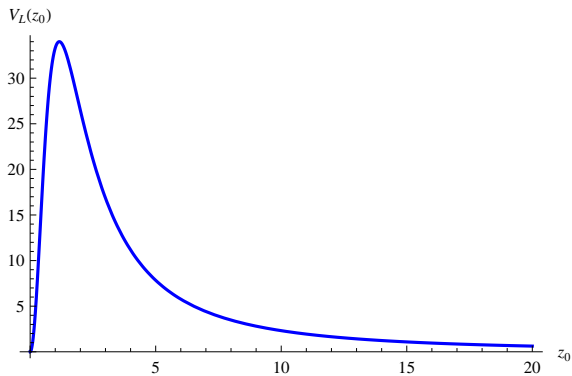}
\end{center}
\caption{(Color online) The shape of the potential of left-hand fermion KK modes $V_{L}(z_0)$ (as the
function of $z_0$). The parameters are set to $k=1,\eta=3,p=1$.}
 \label{figVLz0}
\end{figure*}

Finally, we come to the parameter $z_0$, i.e., the location of the
thin brane. Since the mass of a bound KK mode must satisfy the
relation $m_n\leq \sqrt{V_{L,R}(z_0)}$, the effect of the location
of the thin brane on the number of bound KK modes would be very
important. We plot the shape of $V_{L}(z_0)$ as the function of $z_0$ in Fig. \ref{figVLz0}. It is clear that, for the generalized Yukawa coupling, there is a
maximal value of the potential. Using the first derivative of
$V_{L}(z_0)$ about $z_0$, we obtain that when the formula
\begin{eqnarray}
p=k z_0 \arctan (k z_0)
\end{eqnarray}
is satisfied, the value of $V_{L}(z_0)$ is maximum, which means the
number of the bound KK modes is the most. For the simplest Yukawa
coupling we found the formula is $1=k z_0 \arctan (k z_0)$, meaning
when $k z_0 =1.16$, the value of $V_{L}(z_0)$ is maximum. Here we
present the mass spectra with different $z_0$ for the simplest Yukawa
coupling:
\begin{eqnarray}
\begin{array}{ll}
  m_{Ln}^2 =\{0, 18.19, 30.69\} \cup ~[34.00,\infty) ~~~&\text{for} ~~~~z_0=1.16, \\
  m_{Ln}^2 =\{0, 16.90\} \cup ~[18.57,\infty) ~~~& \text{for}~~~~z_0=0.5,\\
  m_{Ln}^2 =\{0~~~~~~~~\} \cup ~[2.31,\infty) ~~~& \text{for}~~~~z_0=10,\\
\end{array}
\end{eqnarray}
where we set $k=1,\eta=3$. It is clearly that only when $z_0=1.16$,
the number of the bound KK modes is the most. If $z_0 \rightarrow
\infty$, there will be only a zero mode for left-hand fermions, as
$V_{L,R}(z_0)\rightarrow 0$.

\subsubsection{The resonance modes}

Next, we also investigate quasi-localized KK modes of fermions, i.e.,
the fermion resonances. From the shapes of the potentials in
Fig.~\ref{figVLRa} and Fig.~\ref{figVLRb}, we can see that when the
position of the thin brane $z_0$ is not too small and the coupling
constant $\eta$ is large, the potentials $V_{L,R}(z)$ would have a
potential barrier at each side of the location of the thick brane.
Those KK modes with $m^2<V_{L,R}(z_0)$ are bounded and discrete,
while those with $m^2>V_{L,R}(z_0)$ are non-bounded and continuous.
The low-energy continuous modes will generically have very small
amplitudes at the position of the thick brane because they must
tunnel through the potential barrier. Therefore, although low-energy
fermions with $m^2>V_{L,R}(z_0)$ would exist in the theory, their
wave functions will be strongly suppressed on the thick brane, which
leads to a weak coupling with the zero mode and massive bound modes.
However, for higher potential barrier, the KK modes with certain
discrete mass would resonate with the potential and will have larger
probability of being found on the thick brane.

 From (\ref{CoupleEq1a}) and (\ref{CoupleEq1b}) we could conclude
that the masses of the resonances for both chiral fermions are the same
but their parity is reverse, which was discussed in detail in Ref
\cite{Y.X.Liuresonances}. Here we also use the method in that paper to study the
resonances. First, in order to define the parity of the resonances,
we give the initial condition, i.e., $ f(0)=h, f'(0)=0 $ for the
even parity resonances, and $ f(0)=0, f'(0)=h $ for the odd parity
resonances, where $h$ is a constant. Then we use the following
expression to define the relative probability of the KK mode within
a narrow range $-z_b \leq z \leq z_b$ around the thick brane
location \cite{Y.X.Liuresonances}
\begin{equation}
 P_{L,R}(m)=\frac{\int_{-z_b}^{z_b} |f_{L,R}(z)|^2 dz}
                 {\int_{-z_{max}}^{z_{max}} |f_{L,R}(z)|^2 dz},
 \label{Probability}
\end{equation}
where we choose $z_{max}=10 z_b$. The large relative probabilities
would indicate the existence of resonances. Here we give two
examples, for $p=k=1, z_0=5$, $\eta=3$ and $\eta=5$.

\begin{figure*}[htb]
\begin{center}
\includegraphics[width=7.5cm]{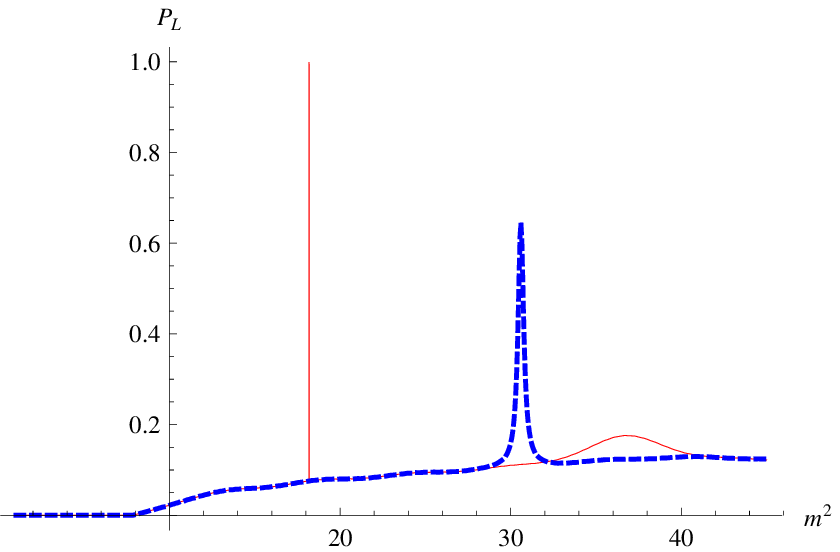}
\includegraphics[width=7.5cm]{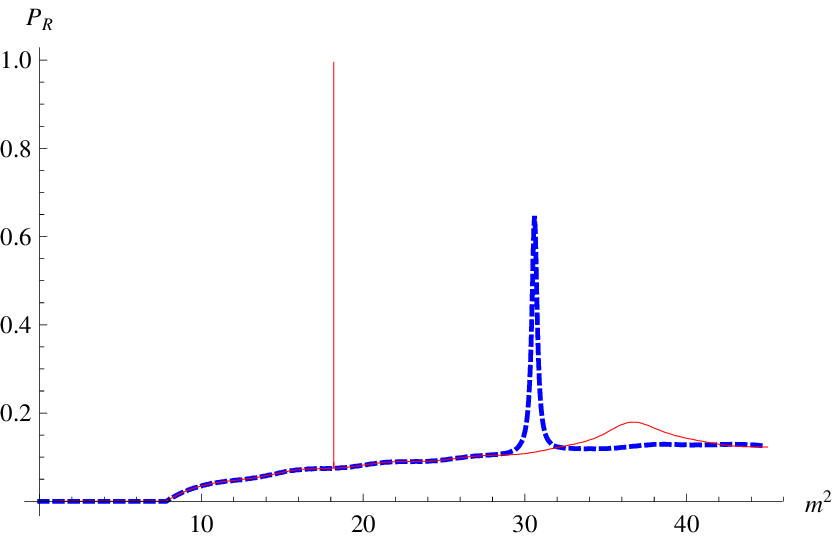}
\end{center}
\caption{(Color online) The probability $P_L(m),P_R(m)$
 for finding the resonances of left and right chiral fermions around the thick brane location.
 The parameters are set to $p=k=1,\eta=3, z_0=5$.
 The red thick lines are for the even resonance,
 and the blue dashing lines are for the odd resonance. }
 \label{figPLPR}
\end{figure*}

\begin{figure*}[htb]
\begin{center}
\includegraphics[width=7.5cm]{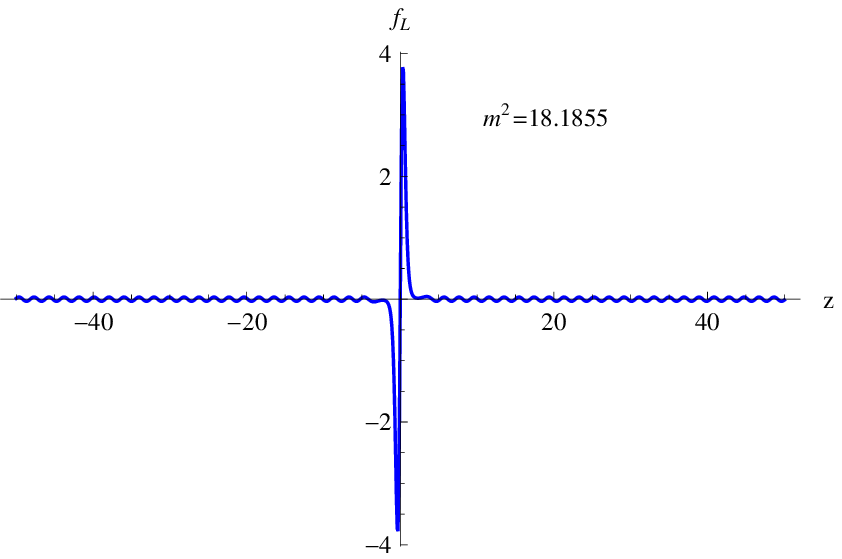}
\includegraphics[width=7.5cm]{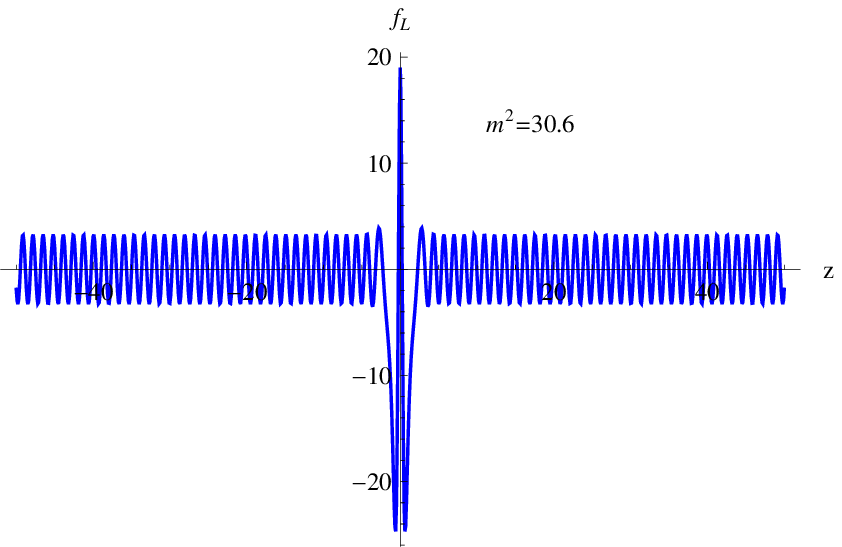}
\end{center}\begin{center}
\includegraphics[width=7.5cm]{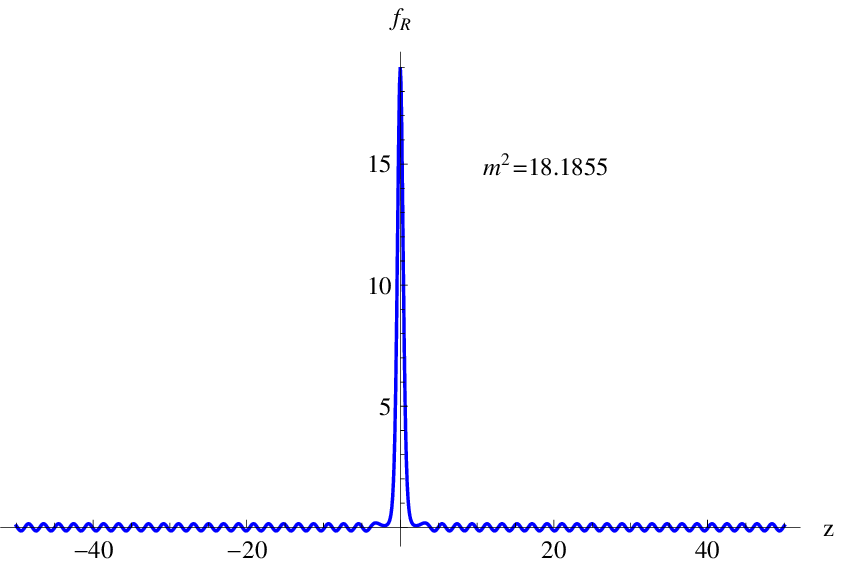}
\includegraphics[width=7.5cm]{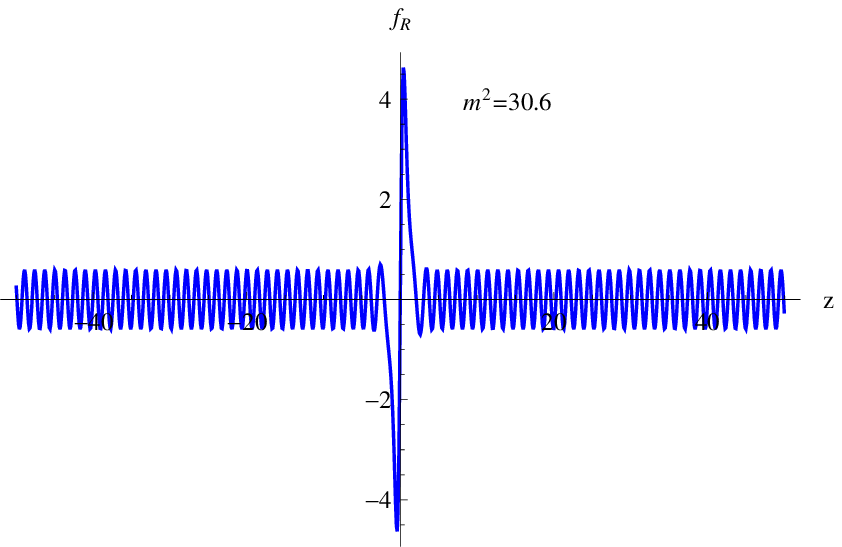}
\end{center}
\caption{(Color online) The shapes of the resonances $f_L$ and $f_R$.
 The parameters are set to $p=k=1,\eta=3, z_0=5$.
 The left pictures correspond to the resonances with $m^2=18.1855$,
 and the right pictures correspond to the resonances with $m^2=30.60$.}
 \label{figFLFR}
\end{figure*}

We found two resonances for both chiral fermions for $p=k=1, z_0=5, \eta=3$, and the corresponding masses are given by $m^2=18.1855$ and $m^2=30.60$. The probabilities and the resonance wave functions are shown in Fig.~\ref{figPLPR} and Fig.~\ref{figFLFR}. For $p=k=1, z_0=5, \eta=5$, there are four resonances for both chiral fermions, of which the masses are $m^2=32.092,~58.93,~79.99,~93.7$, respectively. Comparing with the massive bound KK modes with masses $m^2=32.092,~58.93,~80.00,~93.7$, where the parameters are set $p=k=1, z_0=1.16, \eta=5$, we can see that, when the distance of the two thin branes increases, the discrete massive bound modes would disappear, and the resonances are the remnants of the bound modes. Note that any particle with $m^2>V_{L,R}(z_0)$ produced on the brane can not correspond exactly to a single resonant KK mode because it has a wave function truly localized on the brane \cite{mass5-Dscalar}. It is a wave packet composed of the continuum modes with a Fourier spectrum peaked around one of the resonances. The quasi-localized resonances and fermion tunneling rate were investigated in \cite{RingevalPRD2002}.

\subsection{Spin 0 scalar fields}

In this subsection, let us investigate the localization and mass
spectra of spin 0 scalars on the GRS-inspired brane. The action of a
real scalar field $\Phi$ in 5-dimensional space-time is
\begin{eqnarray}
 S_0 = \int d^5 x  \sqrt{-G}
 \bigg( - \frac{1}{2} G^{M N}\partial_M\Phi \partial_N \Phi
       - V(\Phi,\phi) \bigg), \label{scalarAction}
\end{eqnarray}
where the first part describes the coupling of the scalar field
$\Phi$ with gravity, while the second part $V(\Phi, \phi)$ is the coupling of the
scalar field $\Phi$ with itself and the background scalar field
$\phi$, so it should include $\Phi, \Phi^2, \Phi^3, \Phi^4$ and $(\phi \Phi)^2$ terms, but considering the discrete symmetry we have chosen, $\Phi, \Phi^3$ terms can be eliminated. Note that, since the scalar $\Phi$ is introduced as a perturbation field, high order terms such as $\Phi^6$ can be neglected. We can obtain the equation of motion using the
conformally flat metric (\ref{linee})
\begin{eqnarray}
 \partial_M(\sqrt{-G} G^{M N} \partial_N \Phi)
 + \sqrt{-G} ~U(\phi) \Phi=0, \label{scalarEOM}
\end{eqnarray}
where $U$ is independent of $\Phi$, and is defined as $\frac{\partial V(\Phi,\phi)}{\partial \Phi}=U\Phi+\mathcal {O}(\Phi^2)$ \cite{mass5-Dscalar,GuoAdSBrane}, namely, we mainly consider the $\Phi^2$ and $( \phi \Phi)^2$ terms in $V(\Phi, \phi)$. By
decomposing $\Phi(x,z)=\sum_n\phi_n(x)\chi_n(z)\text{e}^{-3A/2}$, we
find the fifth part of the scalar field $\chi_n(z)$ satisfies the
following Schr\"{o}dinger-like equation
\begin{eqnarray}
  \left[-\partial^2_z+ V_0(z)\right]{\chi}_n(z)
  =m_n^2 {\chi}_n(z),
  \label{SchEqScalar1}
\end{eqnarray}
where we have assumed that the 4-dimensional massive Klein--Gordon
equation
\begin{eqnarray}
 \left(\frac{1}{\sqrt{-\hat{g}}}\partial_\mu(\sqrt{-\hat{g}}
   \hat{g}^{\mu \nu}\partial_\nu) -m_n^2 \right)\phi_n(x)=0
\end{eqnarray}
is satisfied with $\hat{g}^{\mu\nu}=\eta^{\mu\nu}$ the 4-dimensional metric on the brane, and the effective potential is
\begin{eqnarray}
  V_0(z)=\frac{3}{2} A'' + \frac{9}{4}A'^{2}
        + \text{e}^{2A}U(\phi). \label{V0z}
\end{eqnarray}
If we take the scalar potential as the form
\cite{mass5-Dscalar,GuoAdSBrane}
\begin{eqnarray}\label{PotentialVPhiphi}
 V(\Phi,\phi)= \left( \frac{1}{4} \lambda \phi^2
 - \frac{1}{2} u^2\right)\Phi^2 + \mathcal {O}(\Phi^4),
\end{eqnarray}
where $\lambda$ and $u$ are the coupling constants, then $U(\phi)$
can be read as
\begin{eqnarray}
  U(\phi)=\frac{1}{2}\lambda \phi^2 -u^2. \label{PotentialUphi}
\end{eqnarray}
Furthermore, we need the following orthonormality condition to
obtain the 4-dimensional effective action:
\begin{eqnarray}
 \int^{\infty}_{-\infty} dz
 \;\chi_m(z)\chi_n(z)=\delta_{mn}.
 \label{normalizationConditionScalar}
\end{eqnarray}

\subsubsection{Free massless scalars}
We first consider the case without coupling, i.e., 5-dimensional
free massless scalars with $V(\Phi,\phi)=0$. The effective potential
(\ref{V0z}) is
\begin{eqnarray}\label{V0FreeScalar}
  V_0(z)=\left\{\begin{array}{ll}
        \frac{3 k^2 [5(kz)^2-2]}{4 [1 + (kz)^2]^2}, & |z|< z_0 \\
         0, & |z| \geq z_0
      \end{array}~~.\right.
\end{eqnarray}
We note that, if one replaces $k^2$ with $3\lambda$, the above
potential between the two thin branes is just the one given in Ref.
\cite{Liu0708}, where the corresponding potential for scalar KK
modes is $V_0(z)=\frac{9\lambda (15\lambda z^2-2)}{4 (1 + 3\lambda
z^2)^2}$. Hence, if we remove the two thin branes to infinity, we
will recover the same result as in Ref. \cite{Liu0708}. The zero
mode can be solved as
\begin{eqnarray}
 \chi_0(z)
  \propto\left\{\begin{array}{ll}
        \left(1+(k z)^2\right)^{-3/4}, & |z|< z_0 \\
        \left(1+(kz_0)^2\right)^{-3/4}, & |z| \geq z_0
      \end{array}~~.\right.
\end{eqnarray}
Clearly, the normalization condition
(\ref{normalizationConditionScalar}) for the scalar zero mode is not
satisfied for finite $z_0$, which indicates that the 4-dimensional
massless scalar can not be localized between the two thin branes
with a finite distance. In fact, this result can be supported by the
shape of the potential (\ref{V0FreeScalar}): it vanishes beyond the
two thin branes, which indicates that there do not exist any bound
KK modes including the zero mode.

However, there are two limit cases we are interesting in:
$z_0\rightarrow\infty$ and $z_0\ll 1/k$. For the first case, the two
thin branes are infinitely far away from the central thick brane,
and the effective potential (\ref{V0FreeScalar}) is in fact a
volcano type potential. The zero mode $\chi_0(z)$ is normalizable
and is turned out to be
\begin{eqnarray}
 \chi_0(z)= \frac{\sqrt{k}}{\sqrt{2}\left(1+(k z)^2\right)^{3/4}}.
\end{eqnarray}
Namely, when we remove the two thin branes to infinity, we can
localize 4-dimensional massless scalar on the thick brane. In
addition to this massless mode, the potential (\ref{V0FreeScalar})
with $z_0\rightarrow\infty$ suggests that there exists a continuum
gapless spectrum of KK modes with positive $m^2 > 0$.

Now we come to the second limit: $z_0\ll 1/k$, i.e., the distance of
the two thin branes is much less than the thickness of the thick
brane and the thick brane between the two thin branes has a constant
density along the extra dimension. To the lowest order of approximation of $kz_0$, the effective potential
(\ref{V0FreeScalar}) is the known finite square potential well:
\begin{eqnarray}\label{V0FreeScalar0z0}
  V_0(z)=\left\{\begin{array}{cl}
        -\frac{3}{2}k^2 , & |z| < z_0 \\
         0, & |z| \geq z_0
      \end{array}~~.\right.
\end{eqnarray}
Then we would get one and only one bound state with negative
eigenvalue $m_0^2$, which is decided by the relation:
\begin{eqnarray}
 m_0^2&=&-\left(\frac{3}{2}k^2 +m_0^2\right)
       \tan^2\left(\sqrt{\left(\frac{3}{2}k^2
       +m_0^2\right)}\;z_0\right) \nonumber \\
  &\approx& -\frac{9}{4}k^4 z_0^2 .
\end{eqnarray}
Hence, it seems that the system contains a tachyonic state in the limit of $kz_0\ll
1$. However, if one use a more accurate potential (to the next order of $kz_0$), i.e.,
\begin{eqnarray}\label{V0FreeScalar0z0_2}
  V_0(z)=\left\{\begin{array}{cl}
        -\frac{3}{2}k^2 +\frac{27}{4}k^2 z^2 , & |z| < z_0 \\
         0, & |z| \geq z_0
      \end{array}~~\right. ,
\end{eqnarray}
the result will be corrected. From the potential (\ref{V0FreeScalar0z0_2}), we know that the corresponding solution of the bound KK modes is
\begin{eqnarray}\label{WaveFreeScalar0z0_2}
  \chi_n(z)=\left\{\begin{array}{ll}
       c_1 e^{-\frac{1}{2}\sqrt{b}z^2} H_n(b^{1/4}z), & |z| < z_0 \\
       c_2 e^{-m_n z}, & z \geq z_0\\
       c_3 e^{m_n z}, & z \leq -z_0\\
      \end{array}~~\right.,
\end{eqnarray}
where $H_n(z)$ is the Hermite polynomial and $n=(2 m_n^2 + 3k^2 -
3\sqrt{3} k^2)/(6\sqrt{3}k^2)$ is an nonnegative integer. So the
solution of the spectrum $m_n^2$ is
\begin{eqnarray}
  m_n^2 =\frac{3}{2}k^2 (2\sqrt{3}n+\sqrt{3}-1),
\end{eqnarray}
from which we see that the eigenvalue is positive for any
nonnegative $n$, and hence the potential (\ref{V0FreeScalar0z0_2})
has no any bound state with negative eigenvalue, namely, there do
not exist any unstable KK modes .

As for massive KK modes, we can use the approximate potential
(\ref{V0FreeScalar0z0}) to get the resonant spectrum of the scalar KK modes:
\begin{eqnarray}
 m_n^2=\frac{n^2\pi^2}{4z_0^2}-\frac{3k^2}{2}
    \approx \frac{n^2\pi^2}{4z_0^2} ~~~~(n=1,2,3,\cdots).
\end{eqnarray}
It is the same as the
eigenvalue spectrum of the corresponding infinite square potential well.

\subsubsection{Scalars coupled with itself and the domain-wall-generating scalar $\phi$}

Next, we consider the case where the scalar $\Phi$ couples with itself
and the domain-wall-generating scalar $\phi$ with the scalar
potential (\ref{PotentialVPhiphi}). For the GRS-inspired brane
solution, the effective potential (\ref{V0z}) can be expressed as
\begin{eqnarray}\label{VScal}
  V_0(z)=\left\{\begin{array}{ll}
        \frac{21 k^4 z^2}{4(1 + k^2 z^2)^2} + \frac{1}{1 + k^2 z^2}
        \left(6 \lambda \arctan^2(k z) - \frac{3}{2} k^2- u^2\right )
         , & |z| < z_0 \\
         \frac{1}{1 + k^2 z_0^2}
         \left( 6\lambda \arctan^2(k z_0) - {u^2}\right), & |z| \geq z_0
      \end{array}~~.\right.
\end{eqnarray}
We note that the value of the potential at the location of the thick
brane is $V_0(0)=-\frac{3}{2}k^2-u^2$, which is always negative for
nonvanishing $k$. While for $|z|\geq z_0$, the value of the potential is a constant $V_0(z_0)$. So when the relation $6\lambda  \leq u^2 \arctan^{-2}(k z_0)$ is satisfied, there will bound KK modes with negative eigenvalues $m_n^2$ because $V_0(z_0) \leq 0$. Thus in order to get stable solutions, we need to introduce a fine tuning condition $6\lambda = (u^2+v^2)\arctan^{-2}(k z_0)$ with $v$ some constant to be decided by the parameters $u,k,$ and $z_0$, for which $V_0(z_0)\geq v^2/(1+k^2 z_0^2)$ and the potentials become P\"{o}schl-Teller-like potentials. So with the fine tuning condition, there will exist discrete bound KK modes, resonances modes and continues non-bound KK modes with nonnegative eigenvalues and we will get a localized massless scalar on the thick brane. Moreover, in order to localize the massless or massive scalars on the thick brane, the coupling constant $\lambda$ should be large enough.

\begin{figure*}[htb]
\begin{center}
\includegraphics[width=4.5cm]{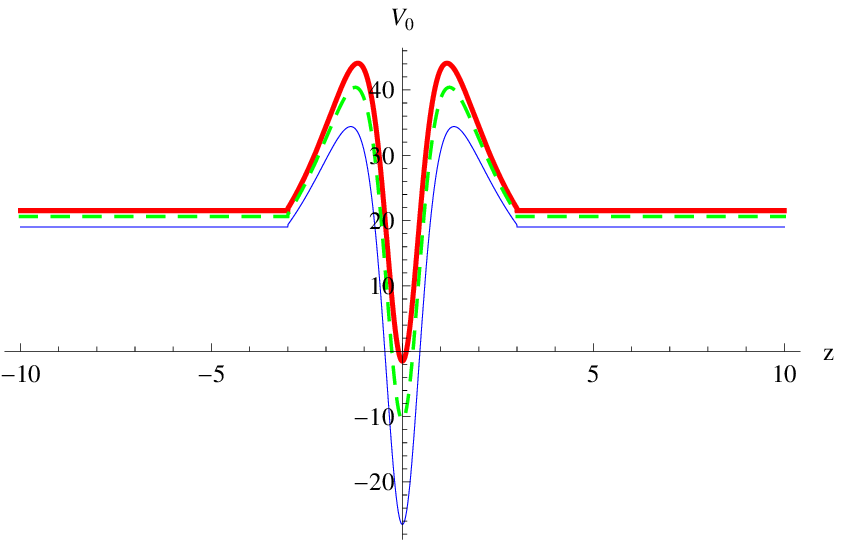}
\includegraphics[width=4.5cm]{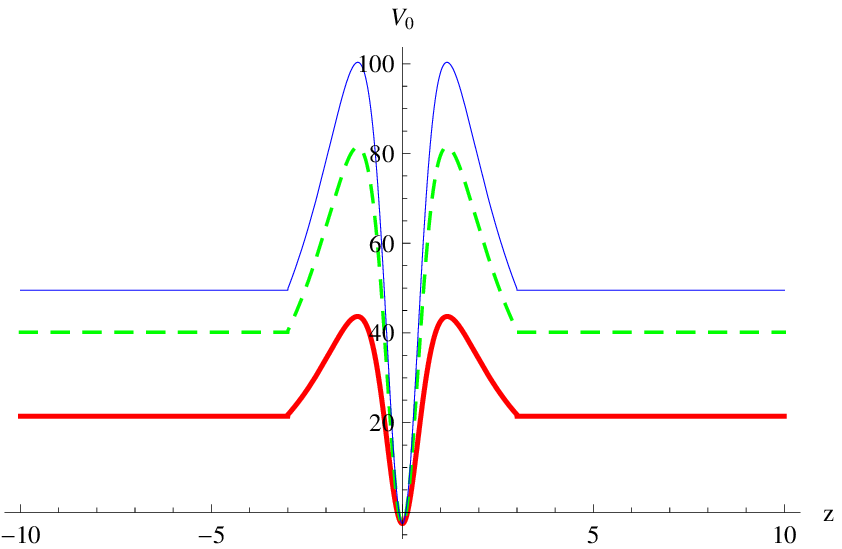}
\includegraphics[width=4.5cm]{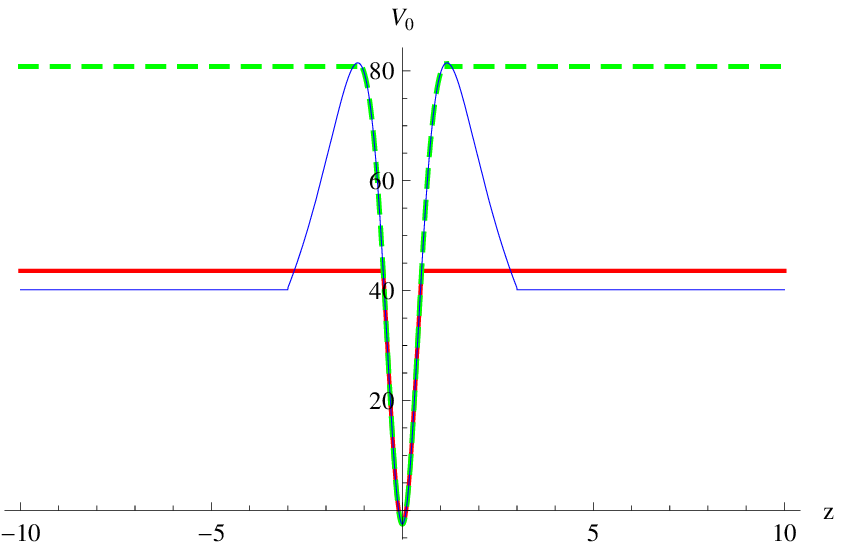}
\end{center}
\caption{(Color online) The shapes of the effective potential $V_0(z)$ of the scalar
KK modes.
 In the left figure, the parameters are set to $k=1,z_0=3,\lambda=23$,
 $u=0$ for red thick lines, $u=3$ for green dashing lines,
 and $u=5$ for blue thin lines.
 In the middle figure, the parameters are set to $k=1,z_0=3,u=1$,
 $\lambda=23$ for red thick lines, $\lambda=43$ for green dashing lines,
 and $\lambda=53$ for blue thin lines.
 In the right figure, the parameters are set to $k=1,\lambda=43,u=1$,
 $z_0=0.5$ for red thick lines, $z_0=1.16$ for green dashing lines,
 and $z_0=3$ for blue thin lines.
 }
 \label{figVSpin0u}
\end{figure*}

The potential is impacted by three parameters, i.e., the coupling constant
$\lambda$, the mass parameter $u$, and the position of the thin brane $z_0$.
The shapes of the potential are shown in Fig.~\ref{figVSpin0u} for different
parameters.

\begin{figure*}[htb]
\begin{center}
\includegraphics[width=7.5cm]{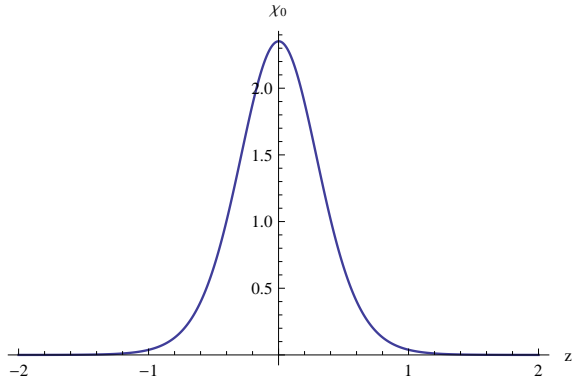}
\end{center}
 \caption{(Color online) The shape of the zero mode $\chi_0$ for the scalar field $\Phi$. The
parameters are set to $k=1, z_0=3, \lambda=23, u=3.1215$.}
 \label{Fermionzeromode}
\end{figure*}

According to the above analysis, we know that there exists a
localized massless scalar on the thick brane if and only if a fine
tuning relation is satisfied. This can be verified by numerical
calculations. As a example, we found that when $k=1, z_0=3,
\lambda=23, u=3.1215$ (the corresponding value of $v$ is $14.3371$),
there exists a massless bound KK mode localized on the thick brane.
The wave function of the zero mode is shown in
Fig.~\ref{Fermionzeromode}.

As for massive bound KK modes, they are mainly impacted by two important
parameters, i.e., the coupling constant $\lambda$ and the position of the thin
brane $z_0$. For simplicity, we set $u=0$ in the following discussion.

\begin{figure*}[htb]
\begin{center}
\includegraphics[width=7.5cm]{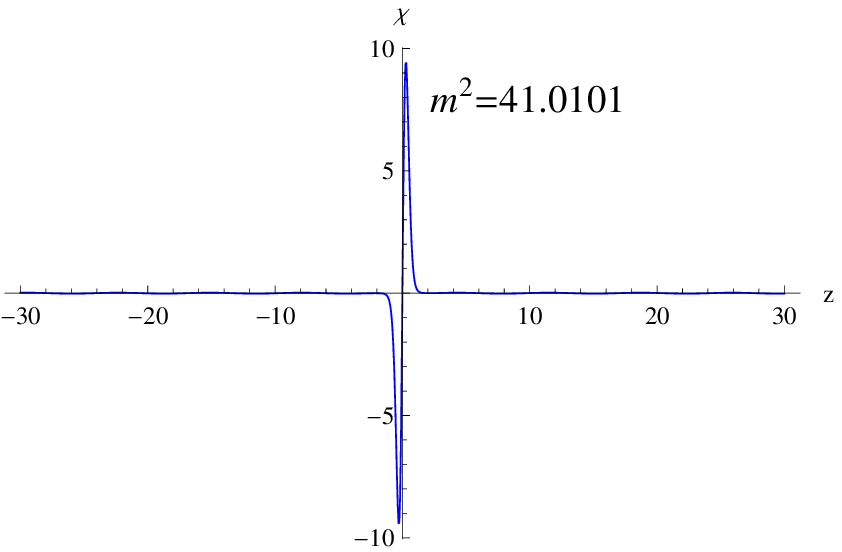}
\includegraphics[width=7.5cm]{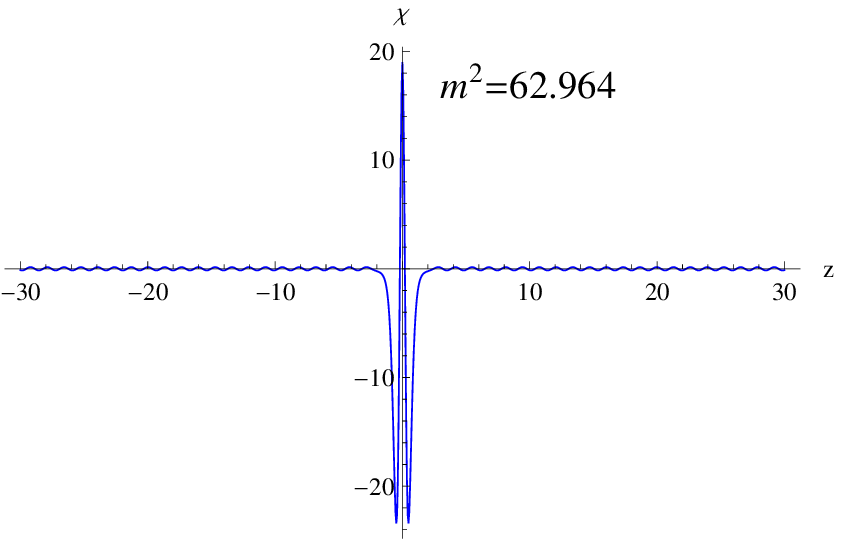}
\includegraphics[width=7.5cm]{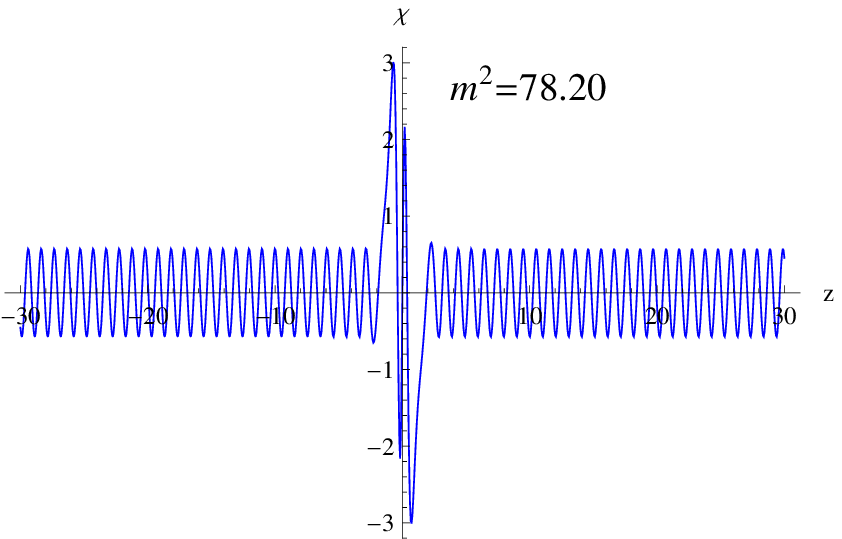}
\end{center}
 \caption{(Color online) The shapes of the scalar resonances. The parameters
are set to $k=1,z_0=3,u=0,\lambda=43$.}
 \label{ScalarResonances1}
\end{figure*}

\begin{figure*}[htb]
\begin{center}
\includegraphics[width=7.5cm]{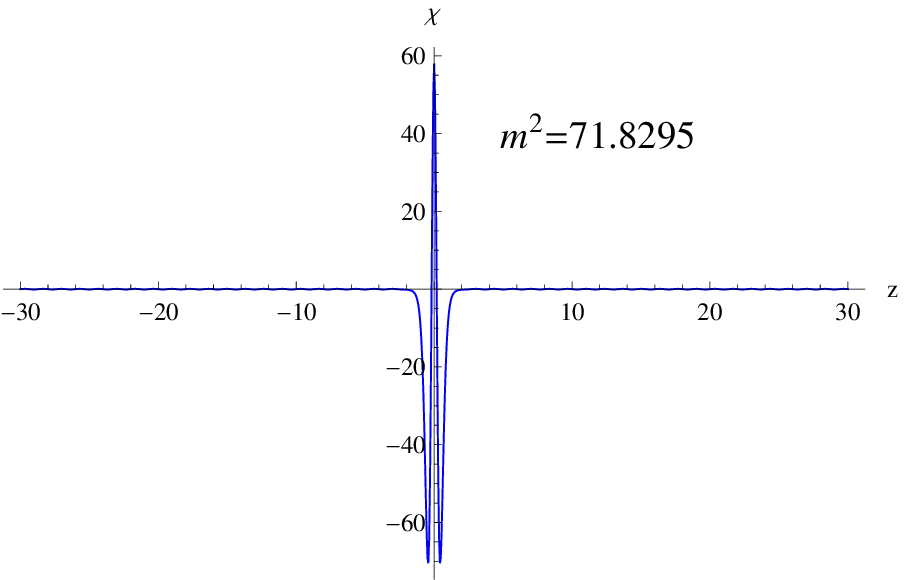}
\includegraphics[width=7.5cm]{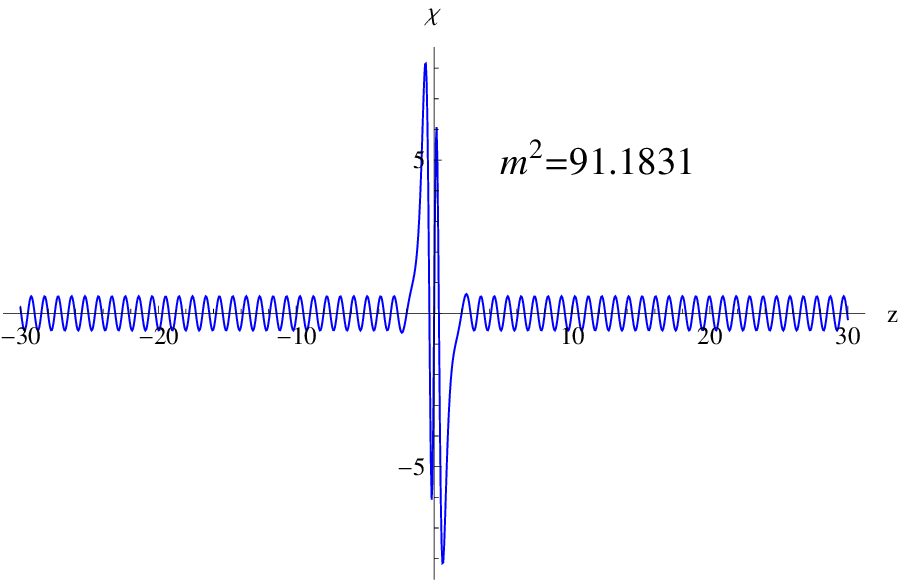}
\end{center}
 \caption{(Color online) The shapes of the scalar resonances. The parameters
are set to $k=1,z_0=3,u=0,\lambda=53$.}
 \label{ScalarResonances2}
\end{figure*}

From the shapes in Fig.~\ref{figVSpin0u}, it is clear that the depth
of both the potential well and the potential barrier increases with
the coupling constant $\lambda$, which means that there may exist
more massive bound KK modes and resonances for stronger coupling.
For example, for $k=1, z_0=3$ and $u=0$, the numerical calculation
shows that, there exists only one massive bound KK mode when
$\lambda=23$ or $\lambda=43$, and exist two massive bound KK modes
when $\lambda=53$. The mass spectra are listed as follows:
\begin{eqnarray}
\begin{array}{ll}
  m_{n}^2 =\{9.31\} \cup ~[21.53,\infty)
      & ~~~\text{for} ~~~\lambda=23, \\
  m_{n}^2 =\{13.54\} \cup ~[40.25,\infty)
     & ~~~\text{for} ~~~\lambda=43, \\
  m_{n}^2 =\{15.29, 46.27\} \cup ~[49.61,\infty)
     & ~~~\text{for} ~~~\lambda=53.
\end{array}
\label{spectraScalar}
\end{eqnarray}
And when $\lambda=23$, there is no resonance, but when $\lambda=43$, there are three
resonances with $m^2=41.0101$, $m^2=62.96$ and $m^2=78.2$. The shapes of them are
shown in Fig.~\ref{ScalarResonances1}. When $\lambda=53$, we found two resonances
with $m^2=71.829$ and $m^2=91.1831$, the shapes of their wave function are shown in
Fig.~\ref{ScalarResonances2}.
\begin{figure*}[htb]
\begin{center}
\includegraphics[width=7.5cm]{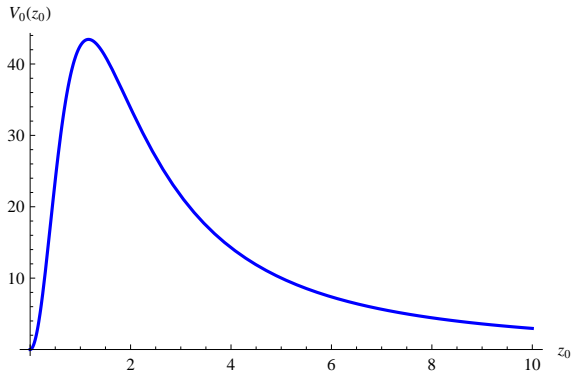}
\end{center}
 \caption{(Color online) The shape of the scalar potential $V_0(z_0)$
 (as the function of $z_0$). The parameters are set to $k=1, \lambda=23$.}
 \label{v-z0Scalar}
\end{figure*}

As shown in Fig.~\ref{figVSpin0u}c, with the increase of
the distance of the two thin branes, the value of the potential
$V_0(z_0)$ at $z\geq z_0$ will first increase and then decrease,
which also can be seen from Fig.~\ref{v-z0Scalar}, where we plot the
shape of $V_0(z_0)$ as a function of $z_0$. The maximum of
$V_0(z_0)$ can be calculated from Eq. (\ref{VScal}). The $z_0$
corresponding to the maximum of $V_0(z_0)$ is the positive root of
the equation $\partial_{z_0} V_0(z_0)=0$, i.e.,
\begin{eqnarray}
 u^2 k z_0 = 6 \lambda \arctan(kz_0) \big(kz_0\arctan(kz_0)-1\big).
\label{z0}
\end{eqnarray}
Especially, for the case of $u=0$, the above equation is reduced to
$1/(kz_0)=\arctan(kz_0)$. The only one solution for this case is
$z_0=1.16234\frac{1}{k}$. This result shows that, when the distance
of the two thin branes is of about the thickness of the thick brane,
the potential that the scalar KK modes feel would be strongest, and
hence there will be the most number of the massive bound KK modes.
For example, the number of the massive bound KK modes with $k=1,
z_0=1.16234$ is more than the case of $k=1, z_0=3$ for $\lambda=23$, or
$\lambda=43$, or $\lambda=53$, which can be seen from the following
mass spectra:
\begin{equation}
\begin{array}{ll}
  m_{n}^2 =\{9.31, 28.24, 41.12\} \cup ~[43.45,\infty), ~~~~~~~~~~~~~\text{for}~~z_0=1.16, \lambda=23,\\
  m_{n}^2 =\{13.56, 41.01, 63.00, 78.30\} \cup ~[81.23,\infty), ~~~\text{for}~~~z_0=1.16, \lambda=43,\\
  m_{n}^2 =\{15.30, 46.27, 71.83, 91.23\} \cup ~[100.12,\infty), ~~\text{for}~~z_0=1.16, \lambda=53.
\end{array}
\label{spectraScalarz0}
\end{equation}
However, when $V_0(z_0)$ reaches the maximum, there will be no
resonances. What interesting is that the eigenvalues of some of
these bound KK modes are almost that of the resonances with some
larger $z_0$. For example, when $k=1, z_0=3, \lambda=43$, there are
three resonances with the eigenvalues $m^2=41.0101$, $m^2=62.96$ and
$m^2=78.2$, and when $k=1, z_0=3, \lambda=53$, there are two
resonances with the eigenvalues  $m^2=71.829$ and $m^2=91.1831$. It
means that, when the distance of the two thin branes increases, some
scalar bound modes would disappear, and the resonances are the
remnants of the bound modes.

\subsection{Spin 1 vector fields}

Lastly, we will investigate the localization of spin 1 vectors.
The action of a 5-dimensional vector field is
\begin{eqnarray}
S_1 = - \frac{1}{4} \int d^5 x \sqrt{-G} G^{M N} G^{R S} F_{MR}
F_{NS}, \label{actionVector}
\end{eqnarray}
where $F_{MN} = \partial_M A_N - \partial_N A_M$. Then using the
background metric (\ref{linee}), the equations of motion of this
system are
\begin{eqnarray}
 \partial_\nu (
      {\eta}^{\nu \rho}\eta^{\mu\lambda}F_{\rho\lambda})
    +{\eta^{\mu\lambda}}e^{-A}\partial_z
      \left(e^{A} F_{4\lambda}\right)  &=& 0, \\
 \eta^{\mu \nu}\partial_\mu F_{\nu 4} &=& 0.
\end{eqnarray}

Next considering the gauge invariant of $\oint A_M dx^M$, we choose $A_4=0$ using gauge freedom. And with this gauge we make a decomposition of the vector field $A_{\mu}(x,z)=\sum_n a^{(n)}_\mu(x)\rho_n(z)e^{-A/2}$, and we can obtain that the KK modes of the vector field satisfy the following Schr\"{o}dinger-like equation
\begin{eqnarray}
  \left[-\partial^2_z +V_1(z) \right]{\rho}_n(z)=m_n^2
  {\rho}_n(z),  \label{SchEqVector1}
\end{eqnarray}
where $m_n$ are the masses of the 4-dimensional vectors, and
$V_1(z)=\frac{A'^2}{4}+\frac{A''}{2}$. Then substitute the Eq. (\ref{SchEqVector1}) into the action (\ref{actionVector}), providing the orthonormality condition
\begin{eqnarray}
 \int^{\infty}_{-\infty} dz \;\rho_m(z)\rho_n(z)=\delta_{mn},
 \label{normalizationConditionVecter}
\end{eqnarray}
we obtain the 4-dimensional effective action:
\begin{eqnarray}
 S_1 &=& \sum_{n}\int d^4 x \sqrt{-\hat{g}}~
       \bigg( - \frac{1}{4}\hat{g}^{\mu\alpha} \hat{g}^{\nu\beta}
             f^{(n)}_{\mu\nu}f^{(n)}_{\alpha\beta}
             - \frac{1}{2}m^2_{n} ~\hat{g}^{\mu\nu}
           a^{(n)}_{\mu}a^{(n)}_{\nu}
       \bigg),
\label{actionVector2}
\end{eqnarray}
where $\hat{g}^{\mu\nu}=\eta^{\mu\nu}$, and $f^{(n)}_{\mu\nu} = \partial_\mu a^{(n)}_\nu - \partial_\nu a^{(n)}_\mu$ is the 4-dimensional field strength tensor.

For the GRS-inspired brane the effective potential of vector KK modes is read as
\begin{eqnarray}\label{Vvector}
  V_1(z)=\left\{\begin{array}{ll}
        \frac{k^2}{1 + k^2 z^2} \big( \frac{5 k^2 z^2}{4(1 + k^2 z^2)}
         - \frac{1}{2} \big ), & |z| < z_0 \\
         0, & |z| \geq z_0
      \end{array}~~.\right.
\end{eqnarray}
The shapes of the potentials are shown below in
Fig.~\ref{figVspin1}. We can obtain the zero mode by solving Eq.
(\ref{SchEqVector1}) with $m^2=0$:
\begin{eqnarray}
 \rho_0(z)
  \propto\left\{\begin{array}{ll}
        \left(1+(k z)^2\right)^{-1/4}, & |z| < z_0 \\
        \left(1+(kz_0)^2\right)^{-1/4}, & |z| \geq z_0
      \end{array}~~.\right.
\end{eqnarray}
For this vector zero mode, the normalization condition
(\ref{normalizationConditionVecter}) can not be satisfied both with a finite $z_0$ or an infinite $z_0$, as the integral $\int\rho_0^2dz$ always tends to infinity. So the massless vector can not be localized on the central thick brane. In fact, this result can be supported by the shape of the potential (\ref{Vvector}): it vanishes
beyond the two thin branes for a finite $z_0$, which indicates that there do not exist any bound KK modes including the zero mode. While for an infinite $z_0$, although the potential becomes a volcano potential, it is not enough to localize the zero mode. It is turned out that the result is the same as in the RS model case. It is known that in the RS model in AdS$_5$ space a spin 1 vector field is not localized neither on a brane with positive tension nor on a brane with
negative tension, so the Dvali-Shifman mechanism should be
considered for the vector field localization \cite{Dvali-Shifman_Mechanism}. In Ref. \cite{Liu0708}, it was also shown that a spin 1 vector can not be trapped on some Weyl thick branes with infinite fifth dimension obtained in Refs.
\cite{ThickBrane1,ThickBrane2,ThickBrane3}. However, it is turned
out that it can be localized on the RS brane in some higher-dimensional cases
\cite{OdaPLB2000113}, or on a thick de Sitter brane \cite{LiuJCAP2009} and on a Weyl thick brane with finite extra dimension \cite{Liu0803}.


\begin{figure*}[htb]
\begin{center}
\subfigure[$z_0=3$]{\label{figVspin1_a}
\includegraphics[width=6.5cm]{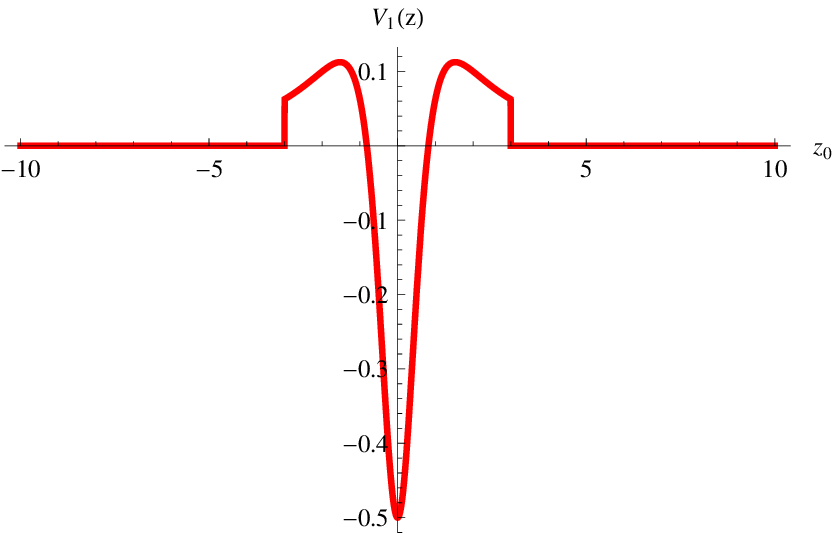}}
\subfigure[$z_0=7$]{\label{figVspin1_b}
\includegraphics[width=6.5cm]{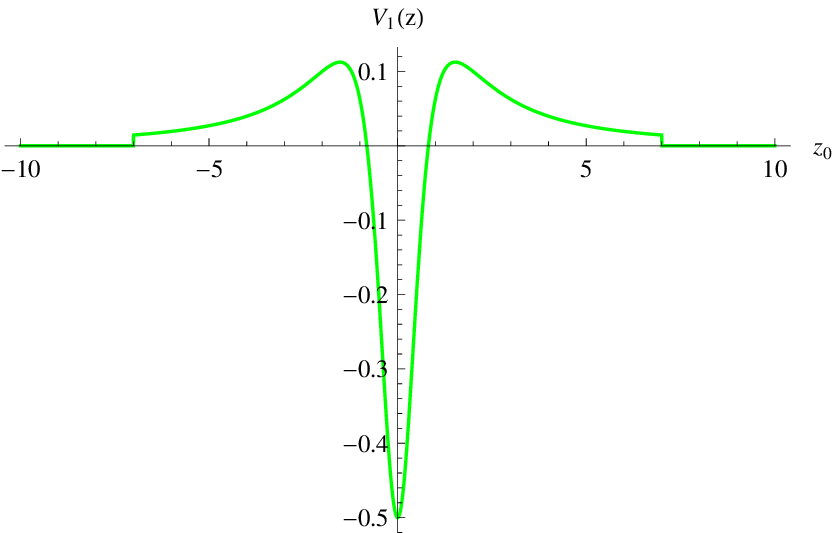}}
\end{center}
 \caption{(Color online) The shapes of potentials of vectors $V_1$.
 The parameters are set to $k=1,z_0=3$ for the thick line,
 and $k=1,z_0=7$ for the thin line.}
 \label{figVspin1}
\end{figure*}

\section{Discussions and conclusions}\label{secConclusion}

In this paper, we have investigated the localization problem and
mass spectra of various bulk matter fields such as spin $\frac{1}{2}$
fermions, spin 0 scalars and spin 1 vectors on the GRS-inspired
brane by presenting the mass-independent potentials of the
KK modes in the corresponding Schr\"{o}dinger equations. In the
braneworld set-up, the extra dimension is infinite and non-compact, and there are one thick brane located at the origin of the extra dimension and two thin branes located at $z=\pm z_0$. The bulk between the two thin branes is AdS$_5$, while
beyond the two thin branes it is Minkowski. Hence, the spectrum for all the fluctuations is made of continue KK modes (for volcano-like potentials) or discrete and continue KK modes (for P\"{o}schl-Teller-like potentials).

For spin $\frac{1}{2}$ fermions the formation of the potentials
(\ref{VL}) and (\ref{VR}) has two sources: the gravity-fermion coupling
$\bar{\Psi} \Gamma^M \omega_M \Psi$ and the scalar-fermion coupling
$\eta \bar{\Psi} F(\phi) \Psi$. If the space-time is flat
($e^A=1$), namely without the gravity-fermion coupling, the
potentials will be {{ P\"{o}schl-Teller-like}} ones for a kink solution, then the massless mode of left  or right chiral fermion can be localized on
the domain wall without any other condition. While without the
scalar-fermion coupling, there will be no bound KK mode because the
potentials vanish. So in order to localize 4-dimensional
fermions, some kinds of scalar-fermion coupling should be
introduced.

In the GRS-inspired model, the potentials for both chiral fermion KK
modes are positive constant for the scalar-fermion coupling with $F(\phi)=\phi^p$  beyond the two thin branes. While between the two thin branes, where the bulk
is AdS$_5$, the minimum values of the potentials for left chiral and right chiral fermions are negative and positive for positive coupling constant
$\eta$, respectively. Hence, we always have massless left chiral fermions on the thick brane for a finite $z_0$ and a positive $\eta$. But for an infinite $z_0$, only when the coupling constant is larger that the critical one, the zero mode for the left chiral fermion can be localized. Furthermore, there exist some discrete bound fermion KK modes and continuous unbound ones. The number of the massive bound KK modes is impacted by three main parameters $\eta, p, z_0$. When the scalar-fermion coupling constant $\eta$ or the coupling strength parameter $p$ increases, the depth of the potential increases, which results in more 4-dimensional massive Dirac fermions trapped on the thick brane.
It was also found that, when the distance of the two thin branes is not
too small and the coupling constant $\eta$ is large, there exist resonance fermions.

For spin 0 scalars $\Phi$, we have considered two cases, the free
massless scalars and the massive scalar coupled with itself and the
background scalar field $\phi$ via a potential $V(\Phi,\phi)=\left(
\frac{1}{4} \lambda \phi^2 - \frac{1}{2} u^2\right)\Phi^2$.
For the free massless scalars, the value of the potential $V_0(z)$
is zero at $|z|\geq z_0$. So, the zero mode can not be localized on the thick brane if the position of the two thin branes $z_0$ is finite. However, if $z_0$ tends to infinity, the zero mode can be localized. When the distance of the two thin branes is much less than the thickness of the thick brane, there are resonances with spectrum $m_n^2\propto n^2$, which is similar to the one of the corresponding infinite square potential well.

For the massive scalars, we need a fine tuning condition to ensure
the localization of the scalar zero mode. With the fine tuning
condition, one can make sure that there is no unstable scalar KK
mode in the model. The number of the massive bound KK modes is
mainly impacted by the parameters $\lambda$ and $z_0$, which is
similar to the case of fermions. With the increase of the coupling constant
$\lambda$, more massive bound KK modes would appear. We also found that there are scalar resonances under some conditions.

For spin 1 vectors, there is no bound KK mode because the effective
potential felt by vectors vanishes outside the two thin branes.

In brief, in this braneworld model with the GRS geometry, the effective Schr\"{o}dinger potentials for the fermions take the form of a modified P\"{o}schl-Teller potential, which in particular has a finite barrier, for the Yukawa and generalized Yukawa couplings to the background kink scalar. So there exist both bound KK modes and resonances, which are different with that for the potentials closer to the standard P\"{o}schl-Teller potentials without the finite barriers in Refs.~\cite{LiuJCAP2009,Liu0803,Liuzhao1004}. 
In fact, it can be seen from (\ref{Vfermion}) that the mass spectrum is decided by three factors: the warp factor $\text{e}^{A}$, the scalar $\phi$, and the coupling $F(\phi)$, namely, the result in the paper is due to a combination of the GRS geometry, the kink like scalar, and the type of couplings to this background that have been chosen.
For the scalars, with the coupling $V(\Phi,\phi)$, the effective Schr\"{o}dinger potential is also modified P\"{o}schl-Teller potential with a finite barrier, which leads to the existence of both the bound KK modes and resonances. While in Refs.~\cite{mass5-Dscalar} and \cite{GuoAdSBrane} with the AdS warped geometry, the Schr\"{o}dinger potential for the scalars with the same coupling $V(\Phi,\phi)$ is a volcano-like potential or an infinite potential.

\section{Acknowledgement}

This work was supported by the Program for New Century Excellent
Talents in University, the National Natural Science Foundation of
China (No. 11075065), the Huo Ying-Dong Education Foundation of
Chinese Ministry of Education (No. 121106), the Doctoral Program Foundation of
Institutions of Higher Education of China (No.
20090211110028), the Natural Science Foundation of Gansu Province,
China (No. 096RJZA055).


\begin{thebibliography}{99}

\bibitem{Rubakov1983}
 V.A. Rubakov and M.E. Shaposhnikov,
    {\em Do we live inside a domain wall?},
    Phys. Lett.  \textbf{B 125} (1983) 136;
  V.A. Rubakov and M.E. Shaposhnikov,
    {\em Extra space-time dimensions: towards a solution to
    {     the cosmological constant problem}},
    Phys. Lett.  \textbf{B 125} (1983) 139.

\bibitem{Akama1983}
 K. Akama,
    {\em Gauge Theory and Gravitation},
    Lect. Notes Phys. \textbf{176} (1983) 267;
 C. Wetterich,
    {\em Chiral fermions in six dimensional gravity},
    Nucl. Phys. \textbf{B 253} (1985) 366;
 S. Randjbar-Daemi and C. Wetterich,
    {\em Kaluza-Klein solutions with noncompact internal spaces},
    Phys. Lett.  \textbf{B 166} (1986) 65.

\bibitem{Antoniadis1990}
 Antoniadis I,
    {\em A possible new dimension at a few TeV},
    Phys. Lett. \textbf{B 246} (1990) 377.

\bibitem{ADD}
 N. Arkani-Hamed, S. Dimopoulos and G. Dvali,
    {\em The hierarchy problem and new dimensions at a millimeter},
    Phys. Lett.  \textbf{B 429} (1998) 263,
    arXiv:hep-ph/9803315;
 I. Antoniadis, N. Arkani-Hamed, S. Dimopoulos and G. Dvali,
    {\em New dimensions at a millimeter to a Fermi and superstrings at a TeV},
    Phys. Lett.  \textbf{B 436} (1998) 257,
    arXiv:hep-ph/9804398.

\bibitem{rs}
 L. Randall and R. Sundrum,
    {\em A Large Mass Hierarchy from a Small Extra Dimension},
    Phys. Rev. Lett. \textbf{ 83} (1999) 3370,
    arxiv:hep-ph/9905221;
 L. Randall and R. Sundrum,
    {\em An alternative to compactification},
    Phys. Rev. Lett. \textbf{ 83} (1999) 4690,
    arXiv:hep-th/9906064.

\bibitem{Lykken}
 J. Lykken and L. Randall,
    {\em The Shape of Gravity},
    JHEP \textbf{0006} (2000) 014,
    arXiv:hep-th/9908076.

\bibitem{GRS}
R. Gregory, V.A. Rubakov and S.M. Sibiryakov,
    {\em  Opening up extra dimensions at ultra-large scales},
    Phys. Rev. Lett. \textbf{84} (2000) 5928,
    arXiv:hep-th/0002072.

\bibitem{GRS2008}
C. Bogdanos, A. Dimitriadis and K.Tamvakis,
      {\em  Synergistic Gravity and the Role of Resonances in GRS-Inspired Braneworlds},
       Class. Quant. Grav. \textbf{25} (2008) 045008,
       arXiv:0706.1015[hep-th].


\bibitem{dewolfe}
 O. DeWolfe, D.Z. Freedman, S.S. Gubser and A. Karch,
    {\em Modeling the fifth dimension with scalars and gravity},
    Phys. Rev. \textbf{D 62} (2000) 046008,
    arXiv:hep-th/9909134.

\bibitem{GremmPLB2000}
 M. Gremm,
    {\em Four-dimensional gravity on a thick domain wall},
    Phys. Lett. \textbf{B 478} (2000) 434,
    arXiv:hep-th/9912060.

\bibitem{gremm}
 M. Gremm,
    {\em Thick domain walls and singular spaces},
    Phys. Rev. \textbf{D 62} (2000) 044017,
    arXiv:hep-th/0002040;
 K. Ghoroku and M. Yahiro,
    {\em Instability of thick brane worlds},
    arXiv:hep-th/0305150;
 A. Kehagias and K. Tamvakis,
    {\em A Self-Tuning Solution of the Cosmological Constant Problem},
    Mod. Phys. Lett. \textbf{ A 17} (2002) 1767,
    arXiv:hep-th/0011006;
 M. Giovannini,
    {\em Gauge-invariant fluctuations of scalar branes},
    Phys. Rev. \textbf{ D 64} (2001) 064023,
    arXiv:hep-th/0106041;
    {\em Localization of metric fluctuations on scalar branes},
    Phys. Rev. \textbf{ D 65} (2002) 064008,
    arXiv:hep-th/0106131;
 S. Kobayashi, K. Koyama and J. Soda,
    {\em Thick brane worlds and their stability},
    Phys. Rev. \textbf{D 65} (2002) 064014,
    arXiv:hep-th/0107025.

\bibitem{Csaki}
 C. Csaki, J. Erlich, T. Hollowood and Y. Shirman,
    {\em Universal Aspects of gravity localized on thick branes},
    Nucl. Phys. \textbf{B 581} (2000) 309,
    arXiv:hep-th/0001033.

\bibitem{CamposPRL2002}
 A. Campos,
    {\em Critical phenomena of thick brane in warped space-time},
    Phys. Rev. Lett. \textbf{88} (2002) 141602,
    arXiv:hep-th/0111207.

\bibitem{varios}
 R. Emparan, R. Gregory and C. Santos,
    {\em Black holes on thick branes},
    Phys. Rev. \textbf{D 63} (2001) 104022,
    arXiv:hep-th/0012100;
 A. Wang,
    {\em Thick de Siter 3branes, dynamic black holes and localization of gravity},
    Phys. Rev. \textbf{D 66} (2002) 024024,
    arXiv:hep-th/0201051;
 A. Melfo, N. Pantoja and A. Skirzewski,
    {\em Thick domain wall space-time with and without reflection symmetry},
    Phys. Rev. \textbf{D 67} (2003) 105003,
    arXiv:gr-qc/0211081;
 K.A. Bronnikov and B.E. Meierovich,
    {\em A general thick brane supported by a scalar field},
    Grav. Cosmol. \textbf{9} (2003) 313,
    arXiv:gr-qc/0402030;
 O. Castillo-Felisola, A. Melfo, N. Pantoja and A. Ramirez,
    {\em Localizing gravity on exotic thick 3-branes},
    Phys. Rev. \textbf{D 70} (2004) 104029,
    arXiv:hep-th/0404083;
 M. Minamitsuji, W. Naylor and M. Sasaki,
    {\em Quantum fluctuations on a thick de Sitter brane},
    Nucl.Phys. \textbf{B 737} (2006) 121,
    arXiv:hep-th/0508093.

\bibitem{Guerrero2002}
 R. Guerrero, A. Melfo and N. Pantoja,
    {\em Self-gravitating domain walls and the thin-wall limit},
    Phys. Rev. \textbf{D 65} (2002) 125010,
    arXiv:gr-qc/0202011.

\bibitem{ThickBraneDzhunushaliev}
 V. Dzhunushaliev, V. Folomeev, D. Singleton and S. Aguilar-Rudametkin,
     {\em 6D thick branes from interacting scalar fields},
     Phys. Rev. \textbf{D 77} (2008) 044006,
     arXiv:hep-th/0703043;
 V. Dzhunushaliev, V. Folomeev, K. Myrzakulov and R. Myrzakulov,
     {\em Thick brane in 7D and 8D spacetimes},
     Gen. Rel. Grav. \textbf{41} (2009) 131,
     arXiv:0705.4014[gr-qc].

\bibitem{ThickBraneBazeia}
 D. Bazeia, F.A. Brito and J.R. Nascimento,
    {\em Supergravity brane worlds and tachyon potentials},
    Phys. Rev. \textbf{D 68} (2003) 085007,
    arXiv:hep-th/0306284;
 D. Bazeia, C. Furtado and A.R. Gomes,
    {\em Brane Structure from a Scalar Field in Warped Spacetime},
    JCAP \textbf{0402} (2004) 002,
    arXiv:hep-th/0308034;
 D. Bazeia, F.A. Brito and A.R. Gomes,
    {\em Locally Localized Gravity and Geometric Transitions},
    JHEP \textbf{0411} (2004) 070,
    arXiv:hep-th/0411088;
 D. Bazeia and A.R. Gomes,
    {\em Bloch Brane},
    JHEP \textbf{0405} (2004) 012,
    arXiv:hep-th/0403141;
 D. Bazeia, F.A. Brito and L. Losano,
    {\em Scalar fields, bent branes, and RG flow },
    JHEP \textbf{0611} (2006) 064,
    arXiv:hep-th/0610233;
 D. Bazeia, A.R. Gomes and L. Losano,
    {\em Gravity localization on thick branes: a numerical approach},
    Int. J. Mod. Phys. \textbf{A 24} (2009) 1135,
    arXiv:0708.3530[hep-th].

\bibitem{ShtanovJCAP2009}
 Y. Shtanov, V. Sahni, A. Shafieloo and A. Toporensky,
    {\em Induced cosmological constant and other features of
    {     asymmetric brane embedding}},
    JCAP \textbf{04} (2009) 023,
    arXiv:0901.3074[gr-qc];
 K. Farakos, N.E. Mavromatos and P. Pasipoularides,
    {\em Asymmetrically Warped Brane Models,
    {    Bulk Photons and Lorentz Invariance}},
    J. Phys. Conf. Ser. \textbf{189} (2009) 012029,
    arXiv:0902.1243[hep-th];
 M. Sarrazin and F. Petit,
    {\em Equivalence between domain-walls and
    {    ``non-commutative'' two-sheeted spacetimes,}
    {    Model-independent matter swapping between branes}},
    Phys. Rev. \textbf{D 81} (2010) 035014,
    arXiv:0903.2498[hep-th];
 V. Dzhunushaliev, V. Folomeev and M. Minamitsuji,
    {\em Thick de Sitter brane solutions in higher dimensions},
    Phys. Rev. \textbf{D 79} (2009) 024001,
    arXiv:0809.4076[gr-qc].

\bibitem{Bazeia0808.2199}
 D. Bazeia, A.R. Gomes, L. Losano and R. Menezes,
    {\em Braneworld Models of Scalar Fields with Generalized Dynamics},
    Phys. Lett. \textbf{B 671} (2009) 402-410,
    arXiv:0808.1815[hep-th].


\bibitem{M. O. Tahim08}
 M.O. Tahim, W.T. Cruz and C.A.S. Almeida,
    {\em Tensor gauge field localization in branes},
    Phys. Rev.\textbf{D 79} (2009) 085022,
    arXiv:0808.2199[hep-th].


\bibitem{0912.1029}
 W.T. Cruz, M.O. Tahim and C.A.S. Almeida,
    {\em Results in Kalb-Ramond field localization and resonances on deformed branes},
    Europhys. Lett. \textbf{88} (2009) 41001,
    arXiv:0912.1029[hep-th].

\bibitem{B. Bajc0302069}
 B. Bajc and G. Gabadadze,
    {\em Massive scalar states localized on a de Sitter brane},
    Phys. Lett. \textbf{D 68} (2003) 064012,
    arXiv:hep-th/0302069.

\bibitem{Angel M. Uranga0208014}
 A.M. Uranga,
    {\em Local models for intersecting brane worlds},
    JHEP \textbf{0212} (2002) 058,
    arXiv:hep-th/0208014.

\bibitem{Ichiro Oda0012013}
 I. Oda,
    {\em Localization of Bulk Fields on AdS$_4$ Brane in AdS$_5$},
    Phys. Lett. \textbf{B 508} (2001) 96,
    arXiv:hep-th/0012013.

\bibitem{BajcPLB2000}
 B. Bajc and G. Gabadadze,
    {\em Localization of matter and cosmological
    {    constant on a brane in anti de Sitter space}},
    Phys. Lett. \textbf{B 474} (2000) 282,
    arXiv:hep-th/9912232.

\bibitem{OdaPLB2000113}
 I. Oda,
    {\em Localization of matters on a string-like defect},
    Phys. Lett. \textbf{B 496} (2000) 113,
    arXiv:hep-th/0006203.


\bibitem{LiuJCAP2009}
  Y.-X. Liu, Z.-H. Zhao, S.-W. Wei and Y.-S. Duan,
    {\em Bulk Matters on Symmetric and Asymmetric de Sitter Thick Branes},
    JCAP \textbf{02} (2009) 003,
    arXiv:0901.0782[hep-th].


\bibitem{Liu0803}
 Y.-X. Liu, L.-D. Zhang, S.-W. Wei and Y.-S. Duan,
    {{\em Localization and Mass Spectrum of Matters on Weyl Thick Branes}},
    JHEP \textbf{0808} (2008) 041,
    arXiv:0803.0098[hep-th].

\bibitem{Liuzhao1004}
 Z.-H. Zhao, Y.-X. Liu, H.-T. Li and Y.-Q. Wang,
   {\em Effects of the variation of mass on fermion localization and resonances on thick branes},
   Phys. Rev. \textbf{D 82} (2010) 084030,
   arXiv:1004.2181[hep-th].

\bibitem{C. A. S. Almeida0901.3543}
 C.A.S. Almeida, R. Casana, M.M. Ferreira Jr and A.R. Gomes,
    {\em  Fermion localization and resonances on two-field thick branes},
    Phys. Rev. \textbf{D 79} (2009) 125022,
    arXiv:0901.3543[hep-th].

\bibitem{Liu0907.0910}
 Y.-X. Liu, C.-E. Fu, L. Zhao and Y.-S. Duan,
    {\em Localization and Mass Spectra of Fermions on Symmetric and Asymmetric Thick Branes},
    Phys. Rev. \textbf{D 80} (2009) 065020,
    arXiv:0907.0910[hep-th].


\bibitem{Alejandra0601161}
 A. Melfo, N. Pantoja and J.D. Tempo,
    {\em Fermion localization on thick branes},
    Phys. Rev. \textbf{D 73} (2006) 044033,
    arXiv:hep-th/0601161.

\bibitem{Ratna0806.0455}
 R. Koley, J. Mitra and S. SenGupta,
    {\em Fermion localization in generalised Randall Sundrum model},
    Phys. Rev. \textbf{D 79} (2009) 041902,
    arXiv:0806.0455[hep-th].

\bibitem{Parameswaran0608074}
 S.L. Parameswaran, S. Randjbar-Daemi and A. Salvio,
    {\em Gauge Fields, Fermions and Mass Gaps in 6D Brane Worlds},
    Nucl. Phys. \textbf{B 767} (2007) 54,
    arXiv:hep-th/0608074.

\bibitem{LiuJHEP2007}
 L. Zhao, Y.-X. Liu and Y.-S. Duan,
    {\em Fermions in gravity and gauge backgrounds on a brane world},
    Mod. Phys. Lett. \textbf{A 23} (2008) 1129,
    arXiv:0709.1520[hep-th].

\bibitem{Mario}
 G. de Pol, H. Singh and M. Tonin,
    {\em Action with manifest duality for maximally
    {     supersymmetric six-dimensional supergravity}},
    Int. J. Mod. Phys. \textbf{A 15} (2000) 4447,
    arXiv:hep-th/0003106.

\bibitem{LiuNPB2007}
 Y.-X. Liu, L. Zhao, X.-H. Zhang and Y.-S. Duan,
    {\em Fermions in Self-dual Vortex Background on a String-like Defect},
    Nucl. Phys. \textbf{B 785} (2007) 234,
    arXiv:0704.2812[hep-th].

\bibitem{LiuVortexFermion}
 Y.-Q. Wang, T.-Y. Si, Y.-X. Liu and Y.-S. Duan,
     {\em Fermionic zero modes in self-dual vortex background},
     Mod. Phys. Lett. \textbf{A 20} (2005) 3045, arXiv:hep-th/0508111;
 Y.-S. Duan, Y.-X. Liu and Y.-Q. Wang,
     {\em Fermionic Zero Modes in Gauge and Gravity Backgrounds on $T^2$},
     Mod. Phys. Lett. \textbf{A 21} (2006) 2019, arXiv:hep-th/0602157;
 Y.-X. Liu, Y.-Q. Wang and Y.-S. Duan,
     {\em Fermionic zero modes in self-dual vortex background on a torus},
     Commun. Theor. Phys. \textbf{48} (2007) 675,
     arXiv:hep-th/0508096.

\bibitem{Rafael200803}
 S. Rafael and S. Torrealba,
    {\em Exact Abelian Higgs Vortices as 6D Brane Worlds},
    arXiv:0803.0313[hep-th].

\bibitem{StojkovicPRD}
 G. Starkman, D. Stojkovic and T. Vachaspati,
    {\em Zero modes of fermions with a general mass matrix},
    Phys. Rev. \textbf{D 65} (2002) 065003,
    arXiv:hep-th/0103039;
    {\em Neutrino zero modes on electroweak strings},
    Phys. Rev. \textbf{D 63} (2001) 085011,
    arXiv:hep-ph/0007071;
 D. Stojkovic,
    {\em Fermionic zero modes on domain walls},
    Phys. Rev. \textbf{D 63} (2000) 025010,
    arXiv:hep-ph/0007343.

\bibitem{ThickBrane1}
 O. Arias, R. Cardenas and I. Quiros,
    {\em Thick Brane Worlds Arising From Pure Geometry},
    Nucl. Phys. \textbf{B 643} (2002) 187,
    arXiv:hep-th/0202130.

\bibitem{ThickBrane2}
 N. Barbosa-Cendejas and A. Herrera-Aguilar,
    {\em 4D gravity localized in non $Z_2$--symmetric thick branes},
    JHEP \textbf{0510} (2005) 101,
    arXiv:hep-th/0511050.

\bibitem{ThickBrane3}
 N. Barbosa-Cendejas and A. Herrera-Aguilar,
    {\em Localization of 4D gravity on pure geometrical thick branes},
    Phys. Rev. \textbf{D 73} (2006) 084022,
    arXiv:hep-th/0603184.


\bibitem{Liu0708}
 Y.-X. Liu, X.-H. Zhang, L.-D. Zhang and Y.-S. Duan,
    {\em Localization of Matters on Pure Geometrical Thick Branes},
    JHEP \textbf{0802} (2008) 067,
    arXiv:0708.0065[hep-th].

\bibitem{Liu0709}
 X.-H. Zhang, Y.-X. Liu and Y.-S. Duan,
    {\em Localization of fermionic fields on braneworlds
    {    with bulk tachyon matter}},
    Mod. Phys. Lett. \textbf{A 23} (2008) 2093,
    arXiv:0709.1888[hep-th].

\bibitem{20082009}
 D. Bazeia, F.A. Brito and R.C. Fonseca,
    {\em Fermion states on domain wall junctions and the flavor number},
    Eur. Phys. J. \textbf{C 63 }(2009) 163,
    arXiv:0809.3048[hep-th];
 P. Koroteev and M. Libanov,
    {\em Spectra of Field Fluctuations in Braneworld Models with Broken Bulk Lorentz Invariance},
    Phys. Rev. \textbf{D 79} (2009) 045023,
    arXiv:0901.4347[hep-th];
 A. Flachi and M. Minamitsuji,
    {\em Field localization on a brane intersection in anti-de Sitter spacetime},
    Phys. Rev. D \textbf{79} (2009) 104021,
    arXiv:0903.0133[hep-th].

\bibitem{ThickBrane4}
 N. Barbosa-Cendejas, A. Herrera-Aguilar, M.A. ReyesSantos and C. Schubert,
    {\em Mass gap for gravity localized on Weyl thick branes},
    Phys. Rev. \textbf{D 77} (2008) 126013,
    arXiv:0709.3552[hep-th];
 N. Barbosa-Cendejas, A. Herrera-Aguilar, U. Nucamendi and I. Quiros,
    {\em Mass hierarchy and mass gap on thick branes with Poincare symmetry},
    arXiv:0712.3098[hep-th].


\bibitem{0803.1458}
 Y. Kodama, K. Kokubu and N. Sawado,
    {\em Localization of massive fermions on the baby-skyrmion branes in 6 dimensions},
    Phys. Rev. \textbf{D 79}, 065024 (2009),
    arXiv:0812.2638[hep-th];
 Y. Brihaye and T. Delsate,
    {\em Remarks on bell-shaped lumps: stability and fermionic modes},
    Phys. Rev. \textbf{D 78} (2008) 025014,
    arXiv:0803.1458[hep-th].

\bibitem{KoleyCQG2005}
 R. Koley and S. Kar,
    {\em Scalar kinks and fermion localisation in warped spacetimes},
    Class. Quantum Grav. \textbf{22} (2005) 753,
    arXiv:hep-th/0407158.

\bibitem{YuXiaoLiu0909.2312}
 Y.-X. Liu, H.-T. Li, Z.-H. Zhao, J.-X. Li and J.-R. Ren,
    {\em Fermion Resonances on Multi-field Thick Branes},
    JHEP \textbf{0910} (2009) 091,
    arXiv:0909.2312[hep-th].


\bibitem{scalarGRS}
 A. Kehagias and K. Tamvakis,
    {\em Localized Gravitons, Gauge Bosons and Chiral Fermions in Smooth Spaces
    {    Generated by a Bounce}},
    Phys. Lett. \textbf{B 504} (2001) 38,
    arXiv:hep-th/0010112.

\bibitem{AdS5}
 P.D. Mannheim,
 \emph{Brane-localized Gravity}, World Scientific Publishing Company, Singapore (2005).

\bibitem{9808016}
  T. Banks, M.R. Douglas, G.T. Horowitz and E. Martinec,
  \emph{AdS Dynamics from Conformal Field Theory},
  arXiv:hep-th/9808016.

\bibitem{9905186}
 I. Bena,
 {\em On the construction of local fields in the bulk of AdS$_5$ and other space},
   Phys. Rev. \textbf{D 62} (2000) 066007,
   arXiv:hep-th/9905186.

\bibitem{Y.X.Liuresonances}
 Y.-X. Liu, J. Yang, Z.-H. Zhao, C.-E. Fu and Y.-S. Duan,
    {{\em Fermion Localization and Resonances on A de Sitter Thick Brane }},
    Phys. Rev. \textbf{D 80} (2009) 065019,
    arXiv:0904.1785[hep-th].

\bibitem{mass5-Dscalar}
 R. Davies and D.P. George,
    {\em Fermions, scalars, and Randall-Sundrum
    {  gravity on domain-wall branes}},
    Phys. Rev. \textbf{D 76} (2007) 104010,
    arXiv:0705.1391[hep-th].


\bibitem{RingevalPRD2002}
  C. Ringeval, P. Peter and J.P. Uzan,
    {\em Localisation of massive fermions on the brane},
    Phys. Rev. \textbf{D 65} (2002) 044016,
    arXiv:hep-th/0109194.

\bibitem{GuoAdSBrane}
  Y.-X. Liu, H. Guo, C.-E. Fu and J.-R. Ren,
    {\em Localization of Matters on Anti-de Sitter Thick Branes},
    JHEP  \textbf{02} (2010) 080,
    arXiv:0907.4424[hep-th].

\bibitem{Dvali-Shifman_Mechanism}
 G.R. Dvali and M.A. Shifman,
    {\em Domain walls in strongly coupled theories},
    Phys. Lett. \textbf{B 396} (1997) 64,
    arXiv:hep-th/9612128.

\end{thebibliography}
\end{document}